\documentclass[prx,twocolumn,english,superscriptaddress,floatfix,longbibliography]{revtex4-2}

\usepackage{graphicx}
\usepackage[T1]{fontenc}
\usepackage{amsmath}
\usepackage{amssymb}
\usepackage{hyperref}
\usepackage{nameref}
\usepackage{tikz}
\usepackage{multirow}
\usepackage{enumitem}
\usepackage{comment}
\usepackage{siunitx}
\begin{document}


\title{SDQC: Distributed Quantum Computing Architecture \\ Utilizing Entangled Ion Qubit Shuttling}

\author{Seunghyun Baek}
\affiliation{Department of Nano Science and Technology \& SKKU Advanced Institute of Nanotechnology (SAINT), Sungkyunkwan University, Suwon, 16419, Korea}

\author{Seok-Hyung Lee}
\email{seokhyunglee@skku.edu}
\affiliation{Department of Quantum Information Engineering, Sungkyunkwan University, Suwon 16419, Korea}
\affiliation{Centre for Engineered Quantum Systems, School of Physics, The University of Sydney, Sydney, New South Wales 2006, Australia}

\author{Dongmoon Min}
\email{dongmoon.min@skku.edu}
\affiliation{Department of Quantum Information Engineering, Sungkyunkwan University, Suwon 16419, Korea}
\affiliation{Department of Computer Science and Engineering, Sungkyunkwan University, Suwon, 16419, Korea}

\author{Junki Kim}
\email{junki.kim.q@skku.edu}
\affiliation{Department of Nano Science and Technology \& SKKU Advanced Institute of Nanotechnology (SAINT), Sungkyunkwan University, Suwon, 16419, Korea}
\affiliation{Department of Quantum Information Engineering, Sungkyunkwan University, Suwon 16419, Korea}

\date{\today}

\begin{abstract}
    We propose Shuttling-based Distributed Quantum Computing (SDQC), a hybrid architecture that combines the strengths of physical qubit shuttling and distributed quantum computing to enable scalable trapped-ion quantum computing.
    SDQC performs non-local quantum operations by distributing entangled ion qubits via deterministic shuttling, combining the high-fidelity and deterministic operations of shuttling-based architectures with the parallelism and pipelining advantages of distributed quantum computing.
    We present (1) a practical architecture incorporating quantum error correction (QEC), (2) pipelining strategies to exploit parallelism in entanglement distribution and measurement, and (3) a performance evaluation in terms of logical error rate and clock speed.
    For a 256-bit elliptic-curve discrete logarithm problem (ECDLP) instance, which requires \num{2871} logical qubits at code distance 13, SDQC achieves a logical error rate which is $1.20^{+0.94}_{-0.45}\times10^{-8}$ of Photonic DQC error rate and $3.79^{+5.09}_{-2.84}\times10^{-3}$ of Quantum Charge-Coupled Device (QCCD) error rate, while providing 2.82 times faster logical clock speed than QCCD.
\end{abstract}

\maketitle

\section{Introduction}
Quantum computing (QC) offers significant computational advantages over classical computing, with exponential speedups for specific problems such as prime factorization \cite{shor_algorithms_1994, shor_polynomial-time_1999} and quantum simulation \cite{cao_quantum_2019}.
These potential benefits have driven extensive efforts to build large-scale quantum systems.
However, as the number of qubits increases, maintaining reliable performance becomes increasingly challenging due to the accumulation of noise and the complexity of control and communication \cite{preskill_quantum_2018}. 
This necessitates the design of scalable quantum architectures, systems that can grow in size, overcoming the compromise between fidelity and speed \cite{kielpinski_architecture_2002, monroe_large-scale_2014}.
Such architectures must also take into account the requirements of quantum error correction (QEC) \cite{shor_scheme_1995, shor_fault-tolerant_1996}, which imposes structural and connectivity constraints on the physical layout \cite{fowler_surface_2012, schwerdt_scalable_2024, lee_ion-trap_2025}.

Trapped ion quantum computing \cite{cirac_quantum_1995} is one among the leading quantum computing platforms, and to realize a scalable architecture in trapped-ion quantum computing, two major approaches have been studied: photonic distributed quantum computing (DQC) \cite{monroe_large-scale_2014} and the quantum charge-coupled device (QCCD) \cite{kielpinski_architecture_2002}.
Photonic DQC enables non-local quantum operations by distributing entanglement pairs through ion-photon interfaces, allowing remote qubits to interact via gate teleportation \cite{gottesman_demonstrating_1999, eisert_optimal_2000, huang_experimental_2004, wan_quantum_2019} or quantum teleportation \cite{bennett_teleporting_1993, bouwmeester_experimental_1997}.
QCCD, in contrast, physically moves ion qubits using dynamically shaped electric potentials, supporting direct quantum state transfer and flexible all-to-all connectivity within a trap \cite{kielpinski_architecture_2002, pino_demonstration_2021, moses_race-track_2023, moses_race-track_2023, ransford2025helios, niroula2025realization, delaney_scalable_2024, lekitsch_blueprint_2017}.
While both approaches extend beyond the limitations of a single ion chain \cite{schiffer_phase_1993, pagano_cryogenic_2018, leung_entangling_2018, landsman_two-qubit_2019, cetina_control_2022}, they differ significantly in terms of communication latency, fidelity, and architectural complexity.

In this study, we propose a hybrid architecture, Shuttling-based Distributed Quantum Computing (SDQC), that combines the advantages of Photonic DQC and QCCD to enable scalable, high-fidelity non-local quantum operations.
SDQC is structured as a multi-core system in which each core operates locally while communicating between cores through entangled ion qubits transported via deterministic shuttling.
While QCCD moves qubits containing quantum data, directly affecting execution time, SDQC moves entanglement pairs without quantum data of quantum circuits during entanglement distribution.
The architecture incorporates inter-core communication protocols, pipelining to exploit parallelism, and a quantum error correction (QEC) model based on color codes \cite{bombin_topological_2006, gidney_new_2023}.
In the architecture assessment, we evaluate SDQC in comparison with Photonic DQC and QCCD in terms of gate execution time and logical error rate.
We further assess application-level performance by estimating execution time and success rate for representative quantum algorithms with the Fermi-Hubbard model and Shor's algorithm.

The remainder of this paper is organized as follows.
Section \ref{section:BackgroundAndMotivation} provides background and motivation, including an overview of trapped-ion quantum computing, its scalability challenges, and the role of QEC.
Section \ref{section:SDQC} introduces SDQC architecture in detail, covering the architectural model, QEC implementation, pipelining strategies, and key properties.
Section \ref{section:Evaluation} presents the evaluation methodology and results, focusing on time cost, logical error rates, and application performance, compared to other reference architectures.

\section{Background and Motivation}{\label{section:BackgroundAndMotivation}}

\begin{figure*}
  \includegraphics[width=0.8\paperwidth]{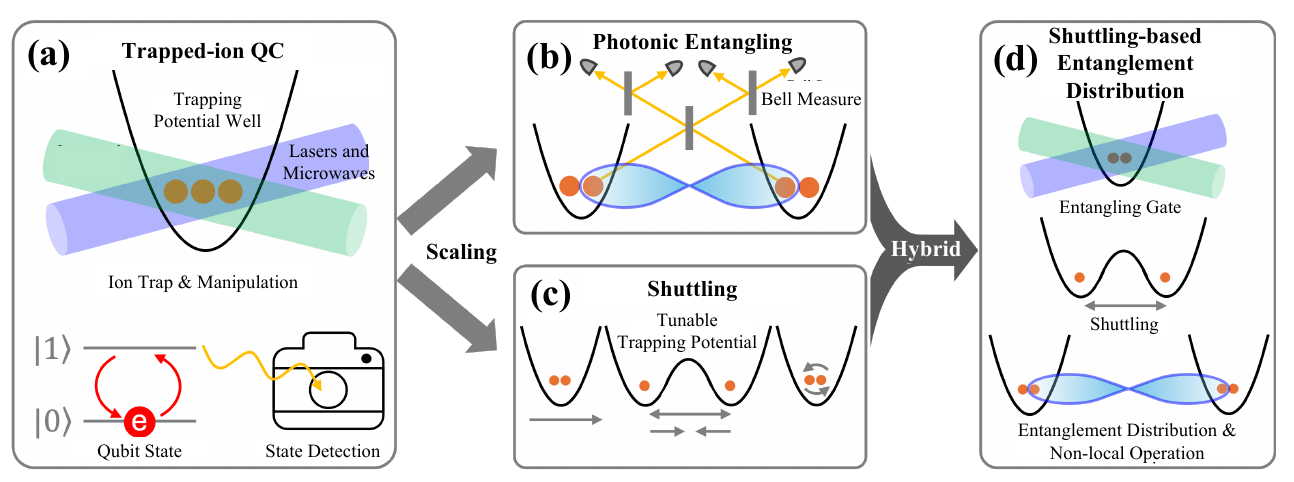}
  \caption{
  Trapped-ion QC and its scalable architectures.
  (a) In trapped-ion QC, ions are trapped in an electric potential well and form a Coulomb crystal that shares a motional mode.
  The quantum information is encoded into the electronic states of ions, which are manipulated by external fields such as lasers and microwaves and measured via state-dependent fluorescence.
  Two main architectures have been proposed for scaling trapped-ion quantum computing: (b) Photonic DQC and (c) QCCD.
  (b) In Photonic DQC, remote entanglement is generated by Bell-state measurements between photons scattered from ions located in separate modules, enabled by ion-photon interfaces. 
  This approach provides asynchronous, modular connectivity with scale-independent distribution latency, while the probabilistic nature and low fidelity of remote entanglement generation remain key challenges.
  (c) QCCD architecture instead achieves scalability by directly shuttling ion qubits by dynamic control of the trapping potential. 
  While it offers deterministic operations and high-fidelity gates, its performance is constrained by scale-dependent latency and ion loss risks during transportation.
  (d) SDQC integrates deterministic ion shuttling with a distributed architecture to enable non-local quantum operations with high fidelity and low latency via asynchronous and pipelined entanglement distribution.}
  {\label{fig:IonTrap}}
\end{figure*}

\subsection{Trapped-ion QC}
Trapped-ion quantum computing utilizes atomic ions as quantum information carriers \cite{leibfried_quantum_2003}, as shown in Fig.~\ref{fig:IonTrap}(a).
The ions are spatially confined in ultra-high vacuum (UHV) environments using electromagnetic potentials, and multiple ions naturally form a linear Coulomb crystal that shares quantized motional modes (phonons) \cite{cirac_quantum_1995, monroe_demonstration_1995, sorensen_quantum_1999}.
The trapping potentials can be finely tuned by adjusting DC and RF voltages applied to the trap electrodes, which are often miniaturized to enable precise spatial control \cite{kaushal_shuttling-based_2020, delaney_scalable_2024}.

Quantum information is encoded in the internal electronic states of atomic ions, which are naturally identical and highly stable.
These well-isolated qubit states minimally interact with the environment, enabling coherence times exceeding one hour \cite{wang_single_2021}.
State discrimination is achieved via state-dependent fluorescence, which allows high-fidelity readout with state preparation and measurement (SPAM) errors as low as $10^{-6}$ \cite{sotirova_high-fidelity_2024}.

Qubit states are manipulated using external control fields either optical or microwave, and entangling operations are mediated by shared phonon modes, with a reported fidelity reaching $99.97(1)~\%$ \cite{loschnauer_scalable_2024, hughes2025trapped}.
Individual optical addressing enables selective control over single ions in a chain, and programmable quantum operations with dozens of ion qubits have been demonstrated across multiple systems \cite{kranzl_controlling_2022, chen_benchmarking_2024}.
Since the phononic mode is shared across all ions in the same chain, trapped-ion systems inherently provide all-to-all connectivity \cite{chen_benchmarking_2024}, in contrast to the limited interaction graphs of other qubit platforms.

Trapped-ion quantum computers exhibit high-fidelity qubit-level operations as well as strong system-level benchmarks \cite{noauthor_cqclquantinuum-hardware-quantum-volume_2025, chen_benchmarking_2024, moses_race-track_2023, pogorelov_compact_2021}.
These capabilities have enabled proof-of-principle demonstrations of quantum error correction\cite{postler2024demonstration, yamamoto2025quantum} and experimental claims of quantum computational advantage over classical computing\cite{liu_certified_2025, kretschmer2025demonstrating}.

\subsection{Scaling trapped-ion QC}
To realize large-scale fault-tolerant quantum computation, approximately a few thousand logical qubits are required, with each logical qubit encoded using many physical qubits \cite{gidney_how_2021}.
In trapped-ion systems, however, the number of physical qubits that can be accommodated in a single linear chain is limited to a few dozens, due to motional mode crowding and associated reduced gate performance \cite{schiffer_phase_1993, pagano_cryogenic_2018, leung_entangling_2018, landsman_two-qubit_2019, cetina_control_2022}.
To address this limitation, two scalable architectural approaches have been proposed: Photonic DQC \cite{monroe_large-scale_2014, hucul_modular_2015, main_distributed_2024, oi_scalable_2006} and QCCD \cite{kielpinski_architecture_2002, pino_demonstration_2021, moses_race-track_2023, delaney_scalable_2024, lekitsch_blueprint_2017, moses_race-track_2023, ransford2025helios, niroula2025realization} architecture.

\subsubsection{Photonic DQC}
Photonic DQC extends trapped-ion QC by distributing and generating entanglement pairs between distant ion qubits using flying photonic qubits as shown in Fig.~\ref{fig:IonTrap}(b) \cite{monroe_large-scale_2014, hucul_modular_2015, main_distributed_2024, oi_scalable_2006}.
In this architecture, each ion chain operates as an independent module, and entangled pairs are generated via ion-photon interfaces that emit photons entangled with ion qubit states \cite{bennett_teleporting_1993, bouwmeester_experimental_1997, huang_experimental_2004, wan_quantum_2019}.
Bell state measurements are performed on photons from different modules using a beamsplitter and coincident detection, which, upon success, heralds entanglement pair generation between the remote ions \cite{simon_robust_2003, moehring_entanglement_2007, cabrillo_creation_1999, hong_measurement_1987, hucul_modular_2015, main_distributed_2024}.
Once entangled, the ion qubits can be used to implement non-local gates through a teleportation-based protocol involving local operation and classical communications (LOCC) \cite{gottesman_demonstrating_1999, eisert_optimal_2000, huang_experimental_2004, wan_quantum_2019, bennett_teleporting_1993, bouwmeester_experimental_1997}.

Photonic DQC approach offers several architectural advantages for scaling trapped-ion qubits.
First, it allows multiple modules to be interconnected, increasing the total number of qubits while enabling long-range connectivity across modules.
These modules interact via distributed entanglements with photonic interfaces, without requiring physical interaction between data qubits in different modules.
Entangled qubits can be prepared asynchronously and remain unentangled with data qubits until interaction, allowing for flexible scheduling in entanglement preparation \cite{bahrani_resource_2025, liu_hardware-software_2025}.
This enables pipelining, which can effectively eliminate the overhead associated with entanglement distribution.

However, realizing Photonic DQC in trapped-ion QC presents substantial experimental challenges.
A primary challenge lies in the implementation of an efficient and reliable quantum interface between ion qubits and photonic qubits \cite{oreilly_fast_2024, ghadimi_scalable_2017, schupp_interface_2021}.
Current approaches typically rely on collecting spontaneously emitted photons via high-numerical-aperture objective lens, which inherently suffer from low success probabilities and limited process fidelity \cite{oreilly_fast_2024}.
Photonic entanglement generation is a heralded process that is intrinsically probabilistic, often requiring tens of thousands of trials for a single successful event.
These constraints result in a slow entanglement generation rate, with reported rates reaching only up to $250~\mathrm{Hz}$ \cite{oreilly_fast_2024}.
Moreover, the bulky and alignment-sensitive nature of current quantum interfaces poses a significant engineering barrier to scaling, particularly in the context of parallelizing remote entanglement generation across multiple modules.

Ion-photon quantum interfaces have been demonstrated using various photonic qubit encoding methods, including frequency \cite{connell_ionphotonic_2021}, polarization \cite{canteri_photon-interfaced_2024}, and time-bin encoding \cite{oreilly_fast_2024, saha_high-fidelity_2025}, with gradual improvements in interface performance.
Cavity quantum electrodynamics (QED) effects have also been employed to enhance ion-photon coupling strength and improve interface efficiency \cite{schupp_interface_2021}.
These developments have enabled preliminary demonstrations of small-scale quantum algorithms based on Photonic DQC architecture \cite{main_distributed_2025}.

\subsubsection{Quantum charge-coupled device (QCCD)}
QCCD architecture achieves scalability by dynamically shuttling ion qubits as shown in Fig.~\ref{fig:IonTrap}(c) \cite{kielpinski_architecture_2002, pino_demonstration_2021, moses_race-track_2023, delaney_scalable_2024, lekitsch_blueprint_2017}.
In QCCD, the spatial configuration of ions is reconfigured to adjust qubit connectivity, enabling large-scale quantum computation without relying on limited interaction topology.
Ion shuttling is implemented by dynamically controlling the voltages applied to segmented DC electrodes, which generate tunable trapping potentials.
This control facilitates various operations, including linear transportation, chain splitting and merging, and positional swapping of ions \cite{kaushal_shuttling-based_2020, pino_demonstration_2021, clark_characterization_2023, delaney_scalable_2024}.

QCCD enables dynamic all-to-all connectivity by physically repositioning ion qubits through shuttling operations, eliminating the need for additional operations (i.e., SWAP gate \cite{baumer_efficient_2024, hashim_optimized_2022} or gate teleportation \cite{gottesman_demonstrating_1999, eisert_optimal_2000, huang_experimental_2004, wan_quantum_2019}) during quantum circuit execution.
As ion shuttling does not disturb qubit states \cite{kaufmann_high-fidelity_2018, kaushal_shuttling-based_2020}, quantum information remains well preserved during transportation, and high-precision shuttling is achieved through accurate electric potential control.
The shuttling operations have a lower error rate than gate operations \cite{wright_reliable_2013, kaufmann_high-fidelity_2018, akhtar_high-fidelity_2023}, which may offer a more reliable approach to long-range connectivity rather than SWAP gate or gate teleportation using photonic entangling.

Although QCCD enables flexible qubit connectivity, several scalability challenges remain.
First, as physical shuttling of ion qubits is required, the overall transportation time increases with system size \cite{webber_efficient_2020, durandau_automated_2023, schoenberger_shuttling_2024}, scaling approximately as the square root of the number of qubits for a two-dimensional QCCD architecture.
While short-distance shuttling can be performed rapidly, large-scale architectures may experience long latency due to longer transportation distances.
As QCCD moves ions containing quantum data (different from DQC moving entanglement pairs), its long transportation time directly affects the operation speed of QCCD.
Second, shuttling-induced motional heating may degrade gate performance, as high-fidelity entangling gates require phonon modes to remain near the ground state \cite{clark_characterization_2023, delaney_scalable_2024}.
The motional heating during shuttling can necessitate re-cooling procedures \cite{barrett_sympathetic_2003, feng_efficient_2020}.
Recent advances in shuttling techniques are addressing this issue by reducing the phonon heating rate during transportation \cite{delaney_scalable_2024}.
Third, simultaneous shuttling of multiple qubits necessitates complex control of electrode voltages, requiring fast and accurate multi-channel analog control \cite{malinowski_how_2023, delaney_scalable_2024}.
Finally, efficient compilation (i.e., scheduling and routing) of shuttling operations for a large number of qubits is computationally challenging.
The optimization of shuttling trajectories and schedules has been shown to be an NP-complete problem \cite{botea_complexity_2021, soeken_boolean_2019, maslov_quantum_2008}, and thus, practical implementations currently rely on heuristic approaches to approximate optimal solutions \cite{webber_efficient_2020, soeken_boolean_2019, maslov_quantum_2008, schoenberger_shuttling_2024}.

Fundamental techniques essential for QCCD architecture, including ion transportation, merging and splitting of ion chains, reliable shuttling across junctions, and optimization of dynamic trapping potentials, have been incrementally demonstrated \cite{pino_demonstration_2021, moses_race-track_2023, delaney_scalable_2024, kaushal_shuttling-based_2020, clark_characterization_2023}.
Based on these developments, experimental implementations of QCCD with dozens of qubits have been realized, enabling the execution of quantum algorithms and system-level benchmarks \cite{moses_race-track_2023, ransford2025helios, niroula2025realization}.
In summary, QCCD architecture provides an alternative scaling path instead of increasing Coulomb crystal size for trapped-ion QC and remains under active development toward large-scale realization.
 
\subsection{Quantum error correction \label{subsec:qec}}
Precise manipulation of quantum information is limited due to its susceptibility to environmental noise and the imperfect fidelity of quantum gate operations.
Furthermore, quantum information collapses when it is measured, thereby prohibiting the backup of quantum data, as stated by the no-cloning theorem \cite{wootters_single_1982}.
To address these challenges, QEC protocols encode logical qubits across multiple physical qubits, introducing redundancy that allows errors to be detected and corrected without directly disturbing the stored quantum information.
Various QEC codes have been proposed, including the Shor code \cite{shor_scheme_1995, shor_fault-tolerant_1996}, surface code \cite{kitaev_fault-tolerant_2003, fowler_surface_2012, horsman_surface_2012}, and color code \cite{bombin_topological_2006, landahl_fault-tolerant_2011, gidney_new_2023, lacroix_scaling_2024}, and are commonly characterized using the notation $[[n,k,d]]$, where $n$ denotes the number of physical qubits, $k$ denotes the number of logical qubits, and $d_\mathrm{code}$ represents the code distance that determines the maximum number of physical errors that can be reliably detected and corrected by the code.

Fault-tolerant quantum computing (FTQC) is achieved by interleaving logical quantum operations with periodic QEC cycles.
In FTQC, a fault-tolerant logical cycle (FT cycle) consists of a logical gate on logical qubits followed by $d_\mathrm{code}$ rounds of QEC cycles.
A typical FTQC workflow includes preparing physical qubits, encoding them into logical qubits, executing a sequence of FT cycles dictated by the target circuit, and finally measuring the logical qubits to obtain computational outcomes.

Each QEC cycle is structured into three stages: syndrome extraction, decoding, and error correction.
In the syndrome extraction stage, specific qubit correlations defined by the QEC code are measured to detect possible deviations from the code space, without directly accessing the encoded quantum information.
Decoding then interprets the syndrome outcomes to infer the most likely error configuration.
Finally, error correction operations are applied to restore the logical qubit state, often implemented virtually through the Pauli frame updates \cite{ryan-anderson_realization_2021}.

The two-dimensional color code \cite{bombin_topological_2006, landahl_fault-tolerant_2011, gidney_new_2023, lacroix_scaling_2024}, which is employed in this study, is a widely used family of QEC codes defined on a two-dimensional trivalent three-colorable lattice, such as a hexagonal lattice.
Each vertex of the lattice hosts a data qubit, and each face supports $X$ and $Z$ stabilizers for syndrome extraction.
The color code enables transversal implementation of all Clifford gates, including the Hadamard, phase, and controlled-NOT (CNOT) gates \cite{kubica_universal_2015, postler_demonstration_2022, ryan-anderson_high-fidelity_2024}, which can be leveraged to achieve highly efficient magic state preparation \cite{chamberland2020very,itogawa2025efficient,gidney2024magic,lee2025low}.
However, decoding the color code involves increased complexity, which results in lower circuit-level error thresholds than those of the surface code.
Nevertheless, there are ongoing efforts to improve decoder performance and scalability, thereby steadily enhancing the viability of the color code as a practical alternative to the surface code \cite{sahay2022decoder,gidney_new_2023, lee_color_2025, koutsioumpas2025colour}.

Motivated by these properties and its growing viability as an alternative to the surface code, we adopt the color code as our primary candidate for realizing FTQC within SDQC architecture. 
In particular, the superdense syndrome extraction circuit for the color code employs Bell pairs as ancilla qubits for stabilizer measurements, which naturally supports segmentation and enables an efficient DQC design (as discussed in detail later). Moreover, the smallest instance of the color code, the $[[7,1,3]]$ Steane code, has already been demonstrated on trapped-ion platforms \cite{ryan-anderson_realization_2021,postler_demonstration_2022}, so our choice of the color code can be viewed as a natural continuation of this experimental trajectory. 
We emphasize, however, that the advantages of SDQC (\textit{e.g.}, reliable, deterministic entanglement distribution for non-local gates and pipelined auxiliary operations) are mainly code-agnostic, and thus we expect similar benefits to be achievable with other codes, including the surface code, although they are not explicitly analyzed in this work.

\section{SDQC: Shuttling-based Distributed Quantum Computing Architecture}{\label{section:SDQC}}
\begin{figure*}
  \includegraphics[width=0.7\paperwidth]{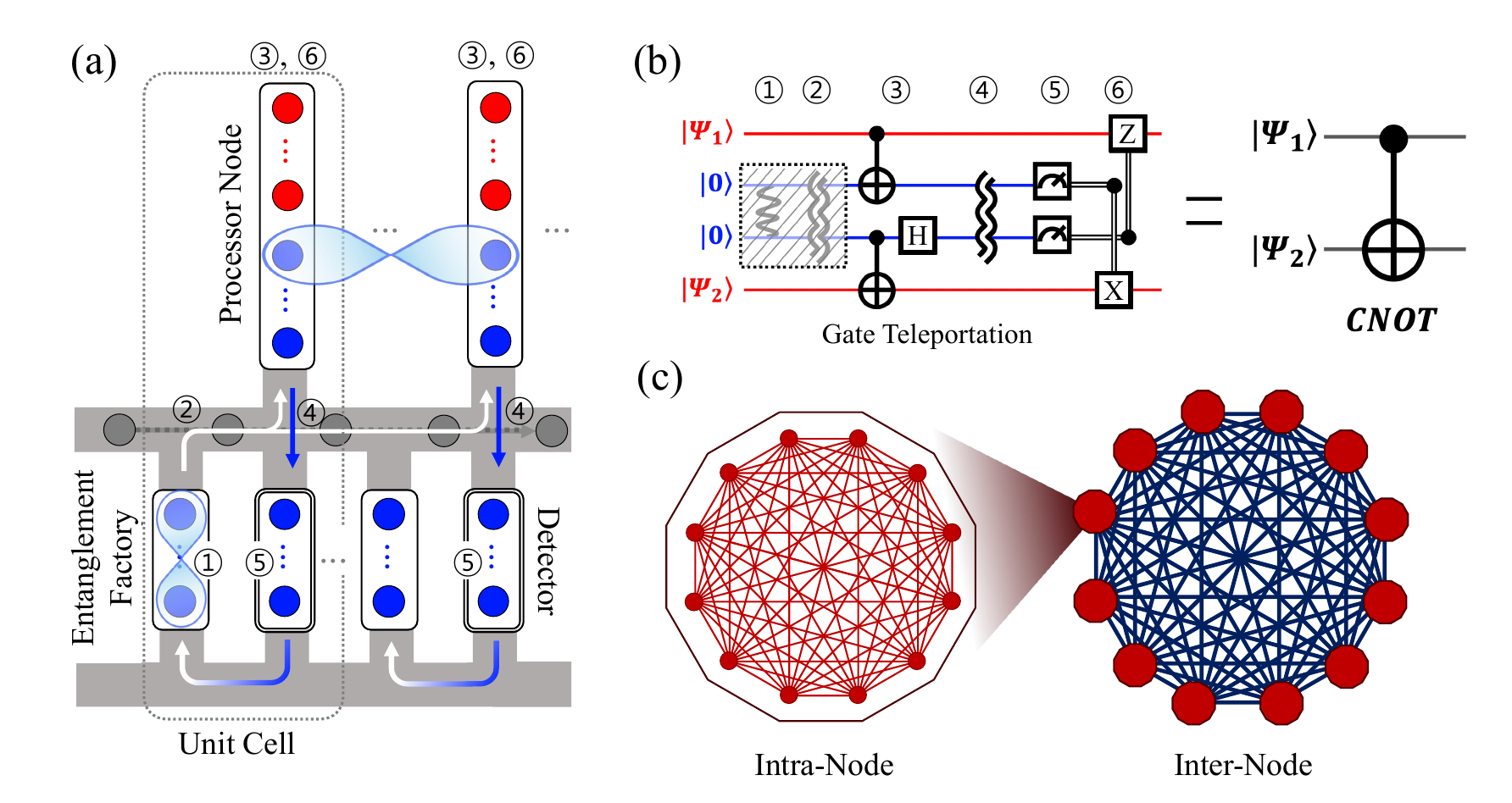}
  \caption{
  Shuttling-based distributed quantum computing (SDQC) architecture.
  (a) Schematic of SDQC architecture.
  Each unit cell consists of a processor node, an entanglement factory, and a detector.
  Unit cells are repeated horizontally and connected via a shuttling network.
  Red and blue circles represent data qubits and entangled qubit pairs for gate teleportation or syndrome extraction (see Sec. \ref{subsection:SyndromeExtractionforErrorCorrection}), respectively.
  Arrows indicate shuttling flow, and numbers correspond to the gate teleportation steps described in (b) and in the main text.
  (b) Inter-node two-qubit gates are implemented via the gate teleportation protocol. 
  An entangled pair is generated and transported from the entanglement factory (Steps~1--2), and they interact with data qubits $\Psi_1$ and $\Psi_2$ individually (Step~3). 
  After the entangled qubits are shuttled to detectors and measured (Steps~4--5), feedforwards are applied to data qubits based on the measurement result.
  The overall result of the protocol results in a remote CNOT between the data qubits.
  Since Steps 1 and 2 do not involve data qubits, their latency can be hidden through pipelining.
  (c) Connectivity topology of SDQC.
  SDQC supports two types of two-qubit gates, intra-node and inter-node gates, resulting in a layered connectivity that reflects both local and non-local interactions. 
}{\label{fig:SDQC}}
\end{figure*}

Here, we introduce our Shuttling-based Distributed Quantum Computing (SDQC) architecture, a hybrid design that combines the complementary advantages of Photonic DQC and QCCD, as shown in Fig.~\ref{fig:IonTrap}(d).
To enable non-local operations at scale, SDQC asynchronously distributes entangled pairs, as in Photonic DQC, using the deterministic and high-fidelity shuttling mechanism of QCCD.
This hybrid approach enables high fidelity and low-latency operations with scale-independent overhead.

In this section, we present a detailed SDQC architecture and describe physical two-qubit gate operations between remote nodes.
We then propose a QEC protocol optimized for the architecture, along with a pipelined execution strategy.
Finally, we discuss circuit compilation and highlight key architectural advantages inherited from both Photonic DQC and QCCD.

\subsection{Overall SDQC architecture}
Figure~\ref{fig:SDQC}(a) illustrates the overall architecture of SDQC.
SDQC comprises horizontally connected unit cells, each containing a processor node, an entanglement factory, and a detector.
These components are interconnected by a shuttling network, which facilitates the transportation of entangled ion qubits.

\textbf{Processor node.}
Each processor node can host multiple ion qubits confined in a single linear chain.
High-fidelity two-qubit gates between arbitrary qubit pairs within a node are enabled by the shared motional mode.
While the number of qubits in each node can vary, an upper limit is imposed to preserve gate performance within a node \cite{schiffer_phase_1993, pagano_cryogenic_2018, leung_entangling_2018, landsman_two-qubit_2019, cetina_control_2022}.
At the beginning of computation, the chain is initialized with data qubits occupying less than half of the available sites, while the remaining sites are left vacant to temporarily accommodate gate teleportation or syndrome extraction qubits.

\textbf{Entanglement factory.} 
An entanglement factory continuously and asynchronously prepares entangled pairs via local phonon-mediated interaction, operating independently of the processor nodes.
These entangled pairs are then transported to processor nodes through the shuttling network and used either for remote two-qubit gates between nodes or for syndrome extraction in QEC cycles (See Section \ref{subsection:SyndromeExtractionforErrorCorrection}).
Each entanglement factory has the same qubit capacity as a processor node.

\textbf{Detector.}
Once the entangled pairs interact with data qubits, they are shuttled to detectors for state measurement.
The measurement outcomes are utilized to apply corrective operations to data qubits, completing either a non-local gate operation or QEC protocol.
Detectors are spatially separated from both processor nodes and entanglement factories to minimize idle qubit decoherence and phonon heating during qubit measurement.
After measurement, entangled pairs are returned to entanglement factories through the bottom recycling path, enabling qubit reuse.

\textbf{Shuttling network.} The shuttling network connects all components inside SDQC (\textit{i.e.}, processor nodes, entanglement factories, and detectors).
Only entangled pairs for gate teleportation or syndrome extraction move through the shuttling network, while data qubits remain stationary in their processor nodes.
To prevent deadlocks caused by bidirectional qubit transportation, the network enforces unidirectional shuttling (horizontal path in Fig.~\ref{fig:SDQC} (a)).
Owing to the distributed nature of DQC, entangled pairs can be generated at any available factory and measured at any detector, regardless of the associated data qubit locations.
This flexibility guarantees that all non-local operations can be executed without requiring bidirectional routing, thereby avoiding transportation conflict and simplifying scheduling.

For more details on architectural assumptions and hardware feasibility, please refer to Appendix \ref{appendix:assumption_sdqc}.

\subsection{Physical two-qubit gate operation}\label{subsection:TQImplementation}
SDQC supports two types of physical two-qubit gates: intra-node and inter-node (non-local/remote) two-qubit gates.
The intra-node two-qubit gate operates within a single ion chain utilizing shared motional modes \cite{cirac_quantum_1995, sorensen_quantum_1999}.
This mechanism allows arbitrary qubit pairs in a chain to interact, enabling all-to-all connectivity within a node.
Recent advanced pulse techniques support not only fast and high-fidelity entangling gates \cite{schafer_fast_2018, nunnerich_fast_2025} but also the parallel execution of multiple two-qubit gates within a node \cite{figgatt_parallel_2019}.

The inter-node two-qubit gate is a non-local quantum operation between processor nodes that relies on gate teleportation using distributed entangled pairs \cite{huang_experimental_2004}.
Figure~\ref{fig:SDQC}(b) shows the process for implementing inter-node two-qubit gates.
SDQC realizes the inter-node CNOT gates with the following steps:

\begin{enumerate}[label=\textbf{Step \arabic*,}, leftmargin=*, itemsep=0pt]
  \item \textbf{Entangled pair generation.} An entangled pair of gate teleportation qubits is generated and prepared in the entanglement factory.

  \item \textbf{Entanglement distribution.} Each gate teleportation qubit of the entangled pair is shuttled to a processor node that hosts the corresponding qubit for the remote two-qubit gate.

  \item \textbf{Local intra-node operation.} Data qubits interact with the delivered gate teleportation qubits by intra-node operations.

  \item \textbf{Shuttling to detectors.} The gate teleportation qubits are shuttled to detectors for measurement.

  \item \textbf{Gate teleportation qubit measurement.} The gate teleportation qubit states are measured in detectors.

  \item \textbf{Feed-forward operations.} Based on the measurement results, corrective operations are applied to the data qubits to complete the non-local gate.
\end{enumerate}

Among these steps, entangled pair generation and distribution (Steps 1 and 2) can be executed in parallel without interfering with other ongoing operations at the processor nodes.
The deterministic shuttling process in SDQC ensures that entangled qubits are delivered precisely when needed, as scheduled during circuit compilation.
This enables the effective hiding of distribution latency, making the performance of inter-node gates independent of the system scale.

While the intra-node gates enable local all-to-all interactions within each chain, inter-node gates extend connectivity across nodes via distributed entanglement. 
This layered structure supports arbitrary qubit pair interactions throughout the entire system, independent of physical qubit placement. 
Furthermore, since all nodes operate independently, entangling gate operations in SDQC can be executed in parallel, enabling scalable non-local gate operations.

Finally, we note that the gate teleportation protocol can be extended to a broader class of operations, such as controlled-unitary gates and multiple CNOTs using additional entangled pairs \cite{eisert_optimal_2000}.
In this work, however, we restrict our focus to remote CNOT gates to simplify the analysis and maintain a consistent evaluation framework.

\subsection{QEC implementation}\label{subsection:SyndromeExtractionforErrorCorrection}

\begin{figure*}
  \includegraphics[width=0.7\paperwidth]{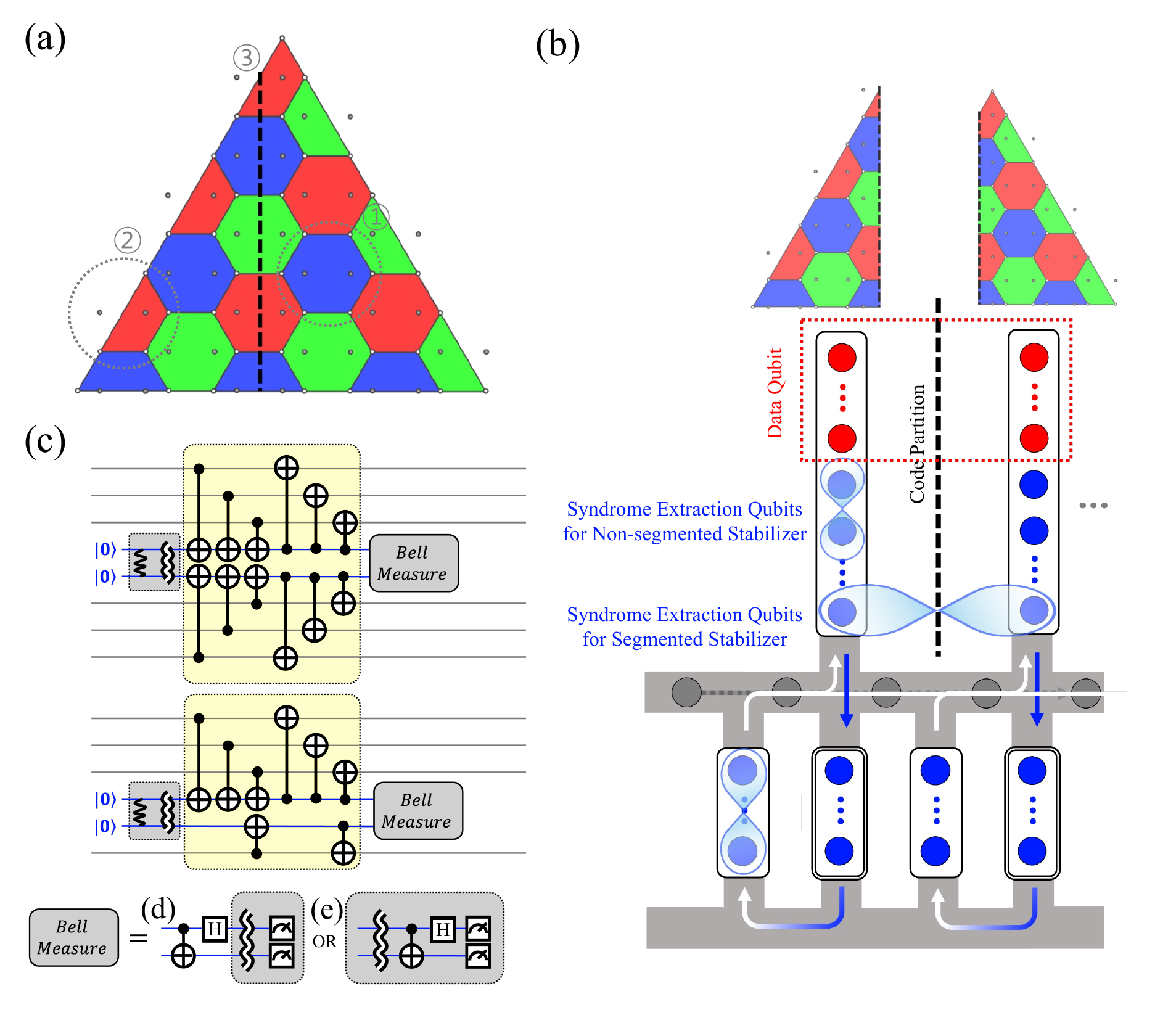}
  \caption{Quantum error correction model.
  (a) Superdense color code.
  Each vertex hosts a data qubit, and the circles on the faces or outside the boundary indicate syndrome extraction qubits.
  There are two types of faces: 1) hexagonal and 2) trapezoidal.
  To overcome the limitation of chain capacity, we place segmented lines such as the dashed line 3), which intersect multiple faces and divide the data qubits into partitions that are located in neighboring chains.
  The stabilizers under the segmentation lines are referred to as segmented stabilizers.
  (b) Physical implementation of syndrome extraction in SDQC.
  The red and blue circles represent data and syndrome extraction qubits, respectively.
  A pair of syndrome extraction qubits for a segmented stabilizer are shuttled to different nodes, interact with data qubits, and then shuttled back to a common detector to be measured in the Bell basis.
  On the other hand, syndrome extraction qubits for non-segmented stabilizers are shuttled to the same node and processed within it, except for the final measurements.
  (c) Superdense syndrome extraction circuit.
  The circuit comprises Bell state preparation and distribution, entangling gates between data and syndrome extraction qubits (yellow), and Bell measurements depicted separately in (d) or (e).
  The upper (lower) circuit measures $X$ and $Z$ stabilizers on a hexagonal (trapezoidal) face.
  Grey and blue lines indicate data and syndrome extraction qubits, respectively.
  Wavy double lines stand for shuttling.
  (d), (e) Bell measurement circuits for non-segmented and segmented stabilizers.
  Gray layers include shuttling and the following operations in detectors.
  }{\label{fig:QECModel}}
\end{figure*}

We propose a detailed QEC implementation optimized for SDQC architecture based on the color code as discussed in Sec.~\ref{subsec:qec}.
As the code distance increases, the number of physical qubits required for a single logical qubit can exceed the capacity of a processor node in SDQC.
In such cases, the color code must be partitioned across multiple nodes, leading to segmented stabilizers whose syndrome extraction demands inter-node communication for entangling gates between syndrome extraction qubits and data qubits.
To mitigate this overhead, we adopt a superdense coding-based syndrome extraction protocol \cite{gidney_new_2023, lacroix_scaling_2024} with distributed entanglement pairs for syndrome extraction, as shown in Fig.~\ref{fig:QECModel}.

In the superdense syndrome extraction protocol, each face requires two syndrome extraction qubits, as illustrated in Fig.~\ref{fig:QECModel}(a).
The $X$ and $Z$ stabilizers on a face are measured via the circuit in Fig.~\ref{fig:QECModel}(c): the two syndrome extraction qubits are first prepared in the Bell state, interacted with data qubits, and finally measured in the Bell basis.

The core of our strategy is to place segmentation lines (\textit{i.e.}, the dashed vertical line in Fig.~\ref{fig:QECModel}(a)) that intersect multiple faces and divide the data qubits into partitions, each residing on a different node.
As described in Fig.~\ref{fig:QECModel}(b), syndrome extraction qubits for segmented stabilizers (supported on the faces crossed by a segmentation line) are distributed to each segment and entangle only with the local subset of data qubits.
This is possible since the part of the superdense circuit that involves data qubits (yellow boxes in Fig.~\ref{fig:QECModel}(c)) is already decomposable into two independent parallel circuits.
After applying the entangling gates, the two syndrome extraction qubits are shuttled to a common detector and jointly measured in the Bell basis [Fig.~\ref{fig:QECModel}(e)].
On the other hand, the process of measuring a non-segmented stabilizer is performed entirely within a single node, except for the initial Bell preparation and the final measurements of the syndrome extraction qubits [Fig.~\ref{fig:QECModel}(d)].
We note that the partition ratio of segmented stabilizers can be chosen among multiple options (\textit{e.g.}, 1:5, 2:4, and 3:3), with the superdense circuit adjusted accordingly.
However, the circuit depth depends on this choice, and it is minimized when the partition is even (\textit{i.e.}, 3:3).

\begin{figure*}
  \includegraphics[width=0.7\paperwidth]{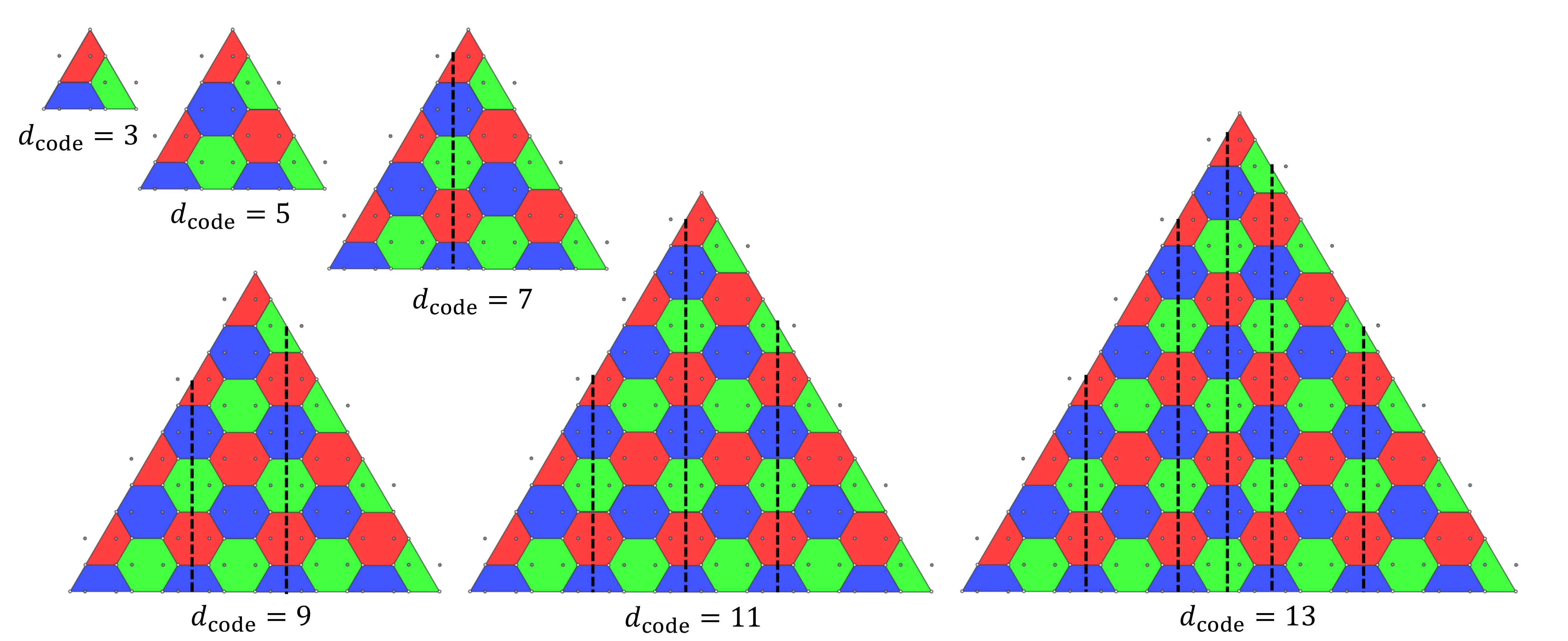}
  \caption{
  Logical qubit segmentation for various code distances.
  Due to the limited qubit capacity of each chain, SDQC segments the physical qubits of a logical qubit across multiple processor nodes or chains to accommodate larger code distances ($d_\mathrm{code} \geq 7$). 
  The dashed lines indicate segmentation boundaries.
  The layout details are provided in Appendix \ref{appendix:PhysicalQubitMappingforLogicalQubit}.
  }{\label{fig:QECMapping}}
\end{figure*}

This approach allows the code distance to be extended without being limited by the qubit capacity of a single processor node.
Figure~\ref{fig:QECMapping} illustrates the code partitioning with code distance ranging from $d_\mathrm{code}=3$ to $d_\mathrm{code}=13$.
Even with the processor node capacity set to 60 qubits in this work, large-distance codes can be implemented without significant overhead.
In addition, the distributed syndrome extraction protocol can be applied to other DQC architectures, provided that entangled pairs for syndrome extraction can be distributed across nodes.
For more details of their physical qubits mapping per logical qubit and QEC sequence, please refer to Appendix \ref{appendix:PhysicalQubitMappingforLogicalQubit} and \ref{appendix:DetailedOperationSequences}.

For fault-tolerant logical gates, SDQC can leverage inter-node operations to realize transversal Clifford gates, including CNOT, Hadamard, and Phase gates, inherently supported by the color code \cite{kubica_universal_2015, postler_demonstration_2022, ryan-anderson_high-fidelity_2024}.
The all-to-all connectivity of inter-node physical two-qubit gates ensures full connectivity of logical qubits.
Furthermore, inter-node operations in SDQC can be executed in parallel, enabling scalable logical gate parallelism.

For non-Clifford gates, including the \textit{T} gate, SDQC requires a magic state factory (\textit{e.g.}, magic state cultivation \cite{gidney2024magic} or its combination with magic state distillation \cite{lee2025low}) to supply distilled magic states for injection.
Once a high-fidelity magic state is available, the injection circuit can be implemented using only Clifford operations and logical qubit measurement, resulting in gate overhead comparable to that of Clifford gates.
As this work represents a foundational step towards achieving a scalable architecture, we focus on the core operations such as transversal gates and syndrome extractions, leaving the explicit integration of magic state factories for future studies.

\subsection{Efficient logical operation protocol}{\label{subsection:Piplining}}
\begin{figure*}
  \includegraphics[width=0.7\paperwidth]{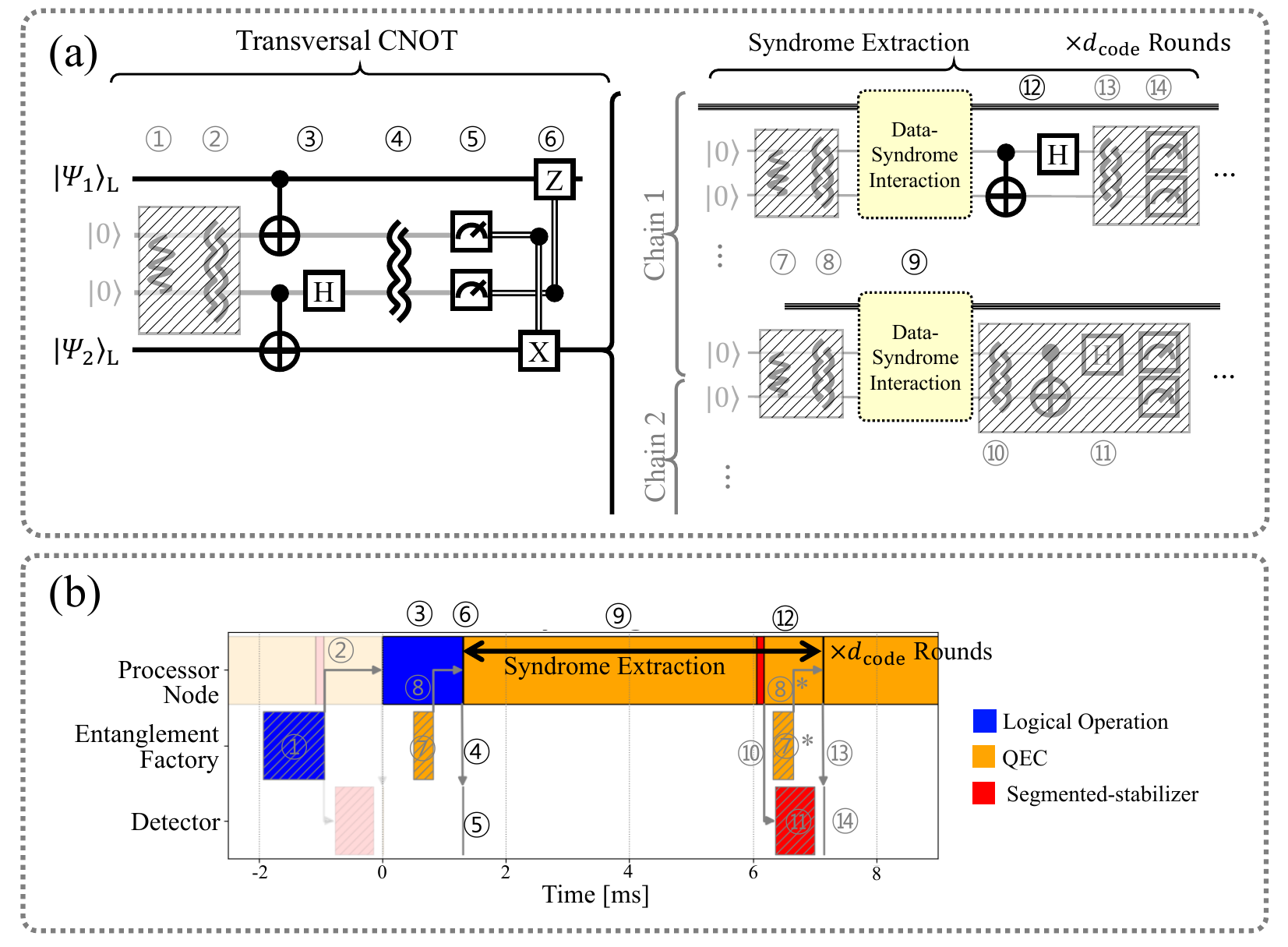}
  \caption{
  Pipelined execution of the FT cycle in SDQC.
  (a) A sequence of a single FT cycle in SDQC, including one logical CNOT gate and repeated syndrome extractions.
  A logical CNOT gate is implemented as transversal gates via parallel gate teleportations (Steps~1--6).
  Following the gate operation, syndrome extractions are repeated for $d_\textrm{code}$ rounds to perform quantum error correction (Steps~7--14).
  All syndromes can be extracted in parallel by leveraging intra-node gates parallelism and appropriate scheduling.  
  Gray-colored steps indicate pipelined stages, while black-colored steps denote operations that effectively set the overall FT cycle latency.
  (b) Temporal diagram of the pipelined FT cycle.
  Entangled pair preparation of gate teleportation qubits (Steps~1--2) and syndrome qubits (Steps~7--8), as well as Bell measurement of syndrome qubits (Steps~10--11, 13-14), are hidden behind data qubit operations (Steps~3--6, 9, 12) by pipelining.
  Blue, orange, and red blocks represent the duration of the logical CNOT gate, QEC cycles, and Bell measurement for segmented-stabilizer, respectively.
  Gray arrows indicate ion shuttling operations across architectural components, where horizontal extent denotes average shuttling duration.
  Timing values in this diagram correspond to a code distance of $d_\textrm{code}=13$
}{\label{fig:SDQCPipelining}}
\end{figure*}

In FTQC, a logical clock time is defined as the execution time for a single FT cycle, including one logical CNOT gate and $d_\mathrm{code}$ rounds of QEC cycles.
To reduce the logical clock time, we propose a pipelined gate execution protocol.
Figure~\ref{fig:SDQCPipelining} shows the pipelined steps and the resulting time cost reduction for the FT cycle.
Specifically, Steps~1--6 in Fig.~\ref{fig:SDQCPipelining}(a) represent the transversal CNOT operation implemented by multiple parallel physical gate teleportation (Sec.~\ref{subsection:TQImplementation}), and Steps~7--14 represent a single QEC cycle for syndrome extraction, which repeats $d_\mathrm{code}$ times.

Within this process, SDQC can pipeline two key stages involving gate teleportation (syndrome extraction) qubits: entangled pairs preparation (Steps~1, 2, 7, and 8) and parallel Bell measurements (steps 11, 13, and 14) as shown in Fig.~\ref{fig:SDQCPipelining}(b). 
The pipelining allows SDQC to hide its latency behind data qubit operations (Steps~3, 6, 9, 10, and 12).
As data qubit operations generally take longer execution time than gate teleportation (syndrome extraction) qubit operations, SDQC can hide non-data-interactive overhead behind the data qubit operation latency.

Implementing such pipelining requires fast and deterministic generation of a large number of entangled pairs.
SDQC is expected to support a sufficient number of entangled pairs even for large code distances, which is feasible with currently available system performance (see Section ~\ref{subsubsection:TimeLogicalTQ}).

\subsection{Compilation}{\label{subsection:Compilation}}
Efficient compilation is essential to achieve optimal performance with QC architectures, although it requires costly processes such as optimal qubit mapping \cite{maslov_quantum_2008}, shuttling path routing \cite{webber_efficient_2020}, and scheduling \cite{schoenberger_shuttling_2024}.
Since compilation process is computational hard \cite{botea_complexity_2021, soeken_boolean_2019, maslov_quantum_2008}, practical approaches rely on heuristic algorithms for approximate solutions \cite{webber_efficient_2020, soeken_boolean_2019, maslov_quantum_2008, schoenberger_shuttling_2024}.

Compilation is particularly challenging in shuttling-based architectures (\textit{e.g.}, QCCD) due to their dynamic qubit mapping.
In QCCD systems, data register allocations change during circuit execution, making optimal quantum circuit mapping highly nontrivial.
In contrast, SDQC shuttles only gate teleportation qubits, while data qubits remain stationary throughout execution.
All data qubits exhibit all-to-all connectivity via gate teleportation, which significantly simplifies compilation.
Moreover, gate teleportation qubits are transferred unidirectionally through the shuttling network, which simplifies the scheduling of shuttling operations.

In Photonic DQC architectures, entangled pairs are generated via probabilistic heralded processes, requiring circuits to wait for successful entanglement events before proceeding.
This induces compilation overhead, as the compiler must handle conditional execution paths and uncertain gate timing.
In contrast, SDQC employs a deterministic entangled pair distribution, ensuring a fixed execution flow without branching or timing uncertainty.
This determinism further simplifies compilation by enabling static scheduling of communication operations.

\subsection{Summary: key properties and advantages of SDQC}
The key idea of designing SDQC is to exploit the complementary advantages of Photonic DQC and QCCD.
From DQC, SDQC inherits the ability to distribute entanglement with flexible timing.
This flexibility enables asynchronous production of entanglement pairs independent of circuit execution applied to data qubits and enables aggressive pipelining, which hides most of the scale-dependent latency associated with entanglement distribution.
Also, the resulting idle time gained from pipelining can be used to improve performance through operations such as entanglement purification, dynamical decoupling, phase shift compensation, and ion-loss recovery. \cite{bennett_purification_1996, wang_single_2021, walther_single_2011, walther_controlling_2012, barrett_sympathetic_2003, pino_demonstration_2021, moses_race-track_2023}.
The latency associated with these supplementary operations can also be largely hidden through the same pipelining mechanism.

On the other hand, the key property inherited from QCCD is the ability to extend the range of deterministic and high-fidelity local operations via qubit shuttling\cite{loschnauer_scalable_2024, wan_quantum_2019}.
A major limitation of Photonic DQC is the performance of the ion-photon interface, which is probabilistic and yields a low generation rate and limited fidelity compared to local gate operations \cite{simon_robust_2003, moehring_entanglement_2007, cabrillo_creation_1999, hong_measurement_1987, hucul_modular_2015, main_distributed_2024}.
SDQC overcomes this limitation by generating and distributing entanglement through local gate operations and shuttling, enabling fast and deterministic delivery of high-fidelity Bell pairs.
Because the fidelity of locally generated entanglement is substantially higher than that of photonic interfaces, shuttling-based distribution significantly improves the fidelity of non-local gates and the overall architectural performance.
The deterministic timing provided by shuttling further enables reliable pipelining by creating predictable idle windows in the execution schedule.

In summary, SDQC integrates high-fidelity entanglement generation with flexible and efficient execution, achieving deterministic operation and scalable performance at practically useful system sizes. 

\section{Evaluation}{\label{section:Evaluation}}
In this section, we evaluate the execution time and error rate of SDQC compared with QCCD and Photonic DQC.
Both metrics are analyzed for physical two-qubit gates and logical two-qubit gates, followed by QEC cycles.
Finally, the architectures are benchmarked by estimating the total execution time and success rate for running quantum applications, specifically the Fermi-Hubbard model simulation \cite{lloyd_universal_1996, jafarizadeh_recipe_2024} and Shor's algorithm for the elliptic-curve discrete logarithm problem (ECDLP) \cite{haner_improved_2020}.

\subsection{Evaluation setup}{\label{subsection:EvaluationSetup}}
\subsubsection{Baseline architectures for comparison}{\label{subsubsection:BaselineArchitecture}}
\begin{figure*}
  \includegraphics[width=0.7\paperwidth]{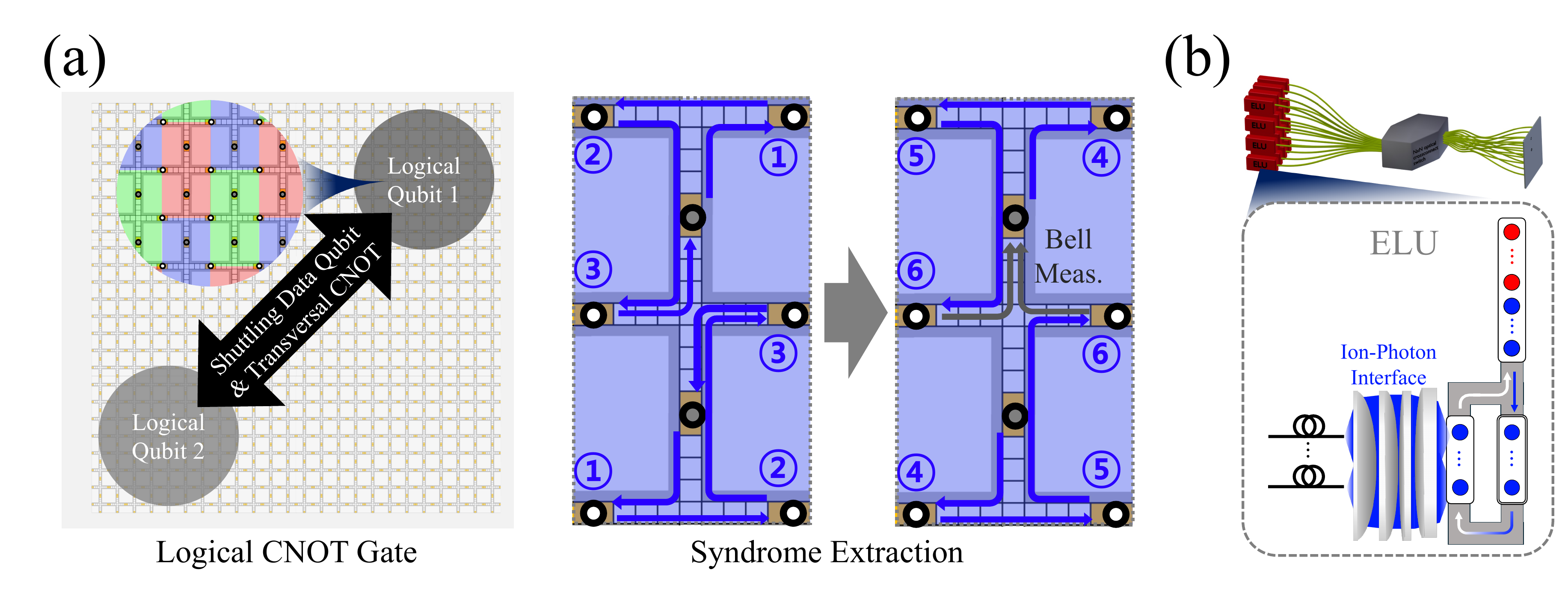}
  \caption{
  Architecture comparison.
  (a) QCCD.
  QCCD has a grid structure, and the color code stabilizers are projected onto the grid.
  The white and gray circles are the data and syndrome extraction qubits, respectively.
  The orange squares under the circles are the operation zones.
  The arrows and Steps~1-6 represent the shuttling routes and entangling gates between data and syndrome extraction qubits for syndrome extraction.
  (b) Photonic DQC adapted from Monroe et al., 2014 \cite{monroe_large-scale_2014}.
  The architecture comprises the elements of a logical unit (ELU) and an optical switch.
  The ELU in this evaluation has a similar structure to SDQC unit cell, and the entanglement factory has an ion-photon interface.}{\label{fig:ComparisonArchitectures}}
\end{figure*}

We evaluate SDQC against the two representative architecture designs: QCCD and Photonic DQC.
For conservative comparison, we model QCCD and Photonic DQC with their optimistic setups.

\textbf{QCCD.} 
Figure~\ref{fig:ComparisonArchitectures}(a) shows the architecture of QCCD.
We assume QCCD as a two-dimensional grid structure following the demonstration of the previous work \cite{delaney_scalable_2024}.
QCCD realizes two-qubit operations of remote qubits by shuttling the target qubit to the location of other qubit and entangling them.

To generate the logical qubits with superdense color code, we assume that QCCD architecture locates the data qubits and syndrome extraction qubits on the transverse and longitudinal paths, respectively, as shown in Fig.~\ref{fig:ComparisonArchitectures}(a).
Each qubit is assumed to have a dedicated operational zone for state preparation, measurement, and logical operation (\textit{i.e.}, orange regions in Fig.~\ref{fig:ComparisonArchitectures}(a)).
QCCD shuttles syndrome extraction qubits to the operational zone of neighboring data qubits and entangles them to run the QEC cycles.
Specifically, following the syndrome extraction process of the Superdense color code, QCCD moves syndrome extraction qubits in the direction of Steps~1--3 while entangling them with the stationary data qubits.
These processes correspond to the $X$ stabilizer of the color code.
Then, QCCD again moves the syndrome extraction qubits in the direction of Steps~4--6, and measures them using Bell-based measurements.
These processes correspond to the $Z$ stabilizer operation of the Superdense color code.

On the other hand, to execute the logical operations among logical qubits, QCCD conducts long-distance shuttling of data qubits to those of another target logical qubit and then transversely entangles them (\textit{i.e.}, Shuttling Data Qubit \& Transversal CNOT in Fig 6(a)).
As the long-distance shuttling requires data qubits to traverse intermediate junctions and qubits, we consider the physical SWAP operation when one qubit faces other qubits during the shuttling.
The shuttling path is optimized by using the routing algorithm proposed in Ref.~\cite{webber_efficient_2020}.

For the conservative comparison for SDQC, we ignore the shuttling overhead of qubits moving back to the origin position after logical two-qubit gate operations. 
This assumption makes the shuttling overhead nearly half compared to the overhead without this assumption.

\textbf{Photonic DQC.} 
Figure~\ref{fig:ComparisonArchitectures}(b) shows the architecture of a Photonic DQC.
Photonic DQC consists of many identical modules, called an elementary logical unit (ELU) \cite{monroe_large-scale_2014}.
We assume the ELU has a similar structure as SDQC's unit cell except for an ion-photon interface, as shown in Fig.~\ref{fig:ComparisonArchitectures}(b).
Each ELU is connected by the ion-photon interface, which generates Bell states between arbitrary ELUs.
By utilizing the generated Bell states between ELUs, Photonic DQC can run QEC cycles and build a logical qubit using many ELUs.
Photonic DQC can also realize remote two-qubit gates and logical two-qubit operations by utilizing the Bell states in a similar manner with SDQC.

For a conservative comparison of SDQC, we assume the following attributes for Photonic DQC.
First, we assume that the entanglement pair (\textit{i.e.}, Bell state) can be generated in a fully parallelized manner without incurring errors to other qubits.
This is an optimistic setup for Photonic DQC because the fully-parallel generation of Bell states has not been demonstrated yet, to the best of our knowledge.
Also, we assume a deterministic generation rate equal to its average value, although the process is inherently probabilistic.
In addition, we assume that the photon delivery error is independent of the delivery distance \cite{collins_experimental_2017, kucera_demonstration_2024}.
Finally, we ignore the latency and loss of optical switches, which connect ELUs by using photonic interconnects.

The detailed QEC implementations for QCCD and Photonic DQC are summarized in Appendix \ref{appendix:DetailedOperationSequences}.

\subsubsection{Number of physical qubits for each architecture}{\label{subsection:QubitSetup}}

We classify the physical qubit count into three categories: qubit count for logical data storage, qubit count for syndrome extraction, and qubit count for gate teleportation.
We derive the required qubit count for each architecture by starting with a baseline superdense color code layout and then adding the additional qubits needed to support the specific QEC cycles and logical operations for each architecture.
The additional qubits belong to the qubit counts for syndrome extraction or gate teleportation.
The physical qubit mapping of the baseline superdense code is summarized in Table~\ref{table:QubitMapping} in Appendix \ref{appendix:PhysicalQubitMappingforLogicalQubit}.
The qubit count for each architecture is determined based on the following rationale.

\textbf{SDQC.} The data qubit count of SDQC is set to be identical to that required for the superdense color code implementation shown in Fig.~\ref{fig:QECMapping}.
For the qubit count of syndrome extraction, we allocate the twice number of syndrome extraction qubits of the baseline superdense color code. 
The doubled allocation is to enable the preparation of Bell states before the end of a QEC cycle, thereby minimizing the time cost by eliminating the delay between QEC cycles.

For the qubit count of gate teleportation, which transports Bell states for logical operations, we calculate the maximum qubit count required to run the target application. The required number of active Bell pairs is derived by multiplying the number of data qubits per logical qubit by the maximum number of logical two-qubit gates within a single application layer.

In addition to these active pairs, the qubit count for gate teleportation also includes spare qubits reserved to mitigate the impact of ion loss.
The operational failure occurs when fewer qubits arrive than the required active qubits after accounting for ion loss events.
We calculate the probability of such failure as a function of spare qubit counts and chose the number of spare qubits to ensure this probability remains below 1~\% of the transversal gate error rate.

\textbf{QCCD.} We set the qubit counts for data and syndrome extraction based on the baseline QCCD structure shown in Fig.~\ref{fig:ComparisonArchitectures}.
For similar reasons to SDQC, we set data qubit count to be the same as those of the superdense color code, and the qubit count for syndrome extraction to twice that of the color code implementation.
We set the number of communication and spare qubits to zero because QCCD directly shuttles data qubits instead of Bell states, and thus does not utilize these qubit types.

\textbf{Photonic DQC } Similar to SDQC, we set the data qubit count to match the color code implementation.
For the qubit count for syndrome extraction, we not only allocate twice the number required by the color code, but also include additional qubits (i.e., the number of `segmented syndrome extraction' in Table~\ref{table:QubitMapping}). 
This additional qubit count is necessary because Photonic DQC requires additional Bell states to enable remote Bell-state measurements between syndrome extraction qubits of segmented stabilizers of different ELUs.
We derive the qubit count for gate teleportation by using the same method as SDQC.
We set the number of spare qubits to zero, as the photonic entanglement process of Photonic DQC has low ion loss probability thanks to its short shuttling distance.

We summarize the detailed equation for calculating the qubit counts in Appendix \ref{appendix:SpaceCostCalculation}.

\subsubsection{Time cost setup}{\label{subsection:TimeCostSetup}}
\begin{table}
    \setlength{\tabcolsep}{6pt}
    \renewcommand{\arraystretch}{1.5}
    \footnotesize{
    \begin{tabular}{lc}
        \hline\hline
        Operation type         & Time [$\mu$s] \\
        \hline
        Single-qubit gate      & 5 \cite{pino_demonstration_2021} \\
        Two-qubit gate         & Max(13.33$N$-54, 100) \cite{leung_robust_2018, murali2020architecting} \\
        Measurement            & 400 \cite{burrell_scalable_2010} \\
        Cooling                & 300 \cite{feng_efficient_2020} \\
        Photonic entangling    & 4000 \cite{oreilly_fast_2024} \\
        Stable transportation  & 46.9 (for 375 $\mu$m) \cite{delaney_scalable_2024} \\
        Fast transportation    & 4.6 (for 375 $\mu$m) \cite{clark_characterization_2023} \\
        Split                  & 128 \cite{pino_demonstration_2021} \\
        Merge                  & 128 \cite{pino_demonstration_2021} \\
        Physical swap          & 200 \cite{pino_demonstration_2021} \\
        \hline\hline
    \end{tabular}}
    \caption{Time Cost of Unit Operation. $N$ denotes the number of physical qubits in a chain.}
    \label{table:UnitOperationTime}
\end{table}
Table~\ref{table:UnitOperationTime} shows the time cost of unit operations.
We select the values from the previously demonstrated works, showing a reasonable scale of latency.
For two-qubit gate, we model its latency as linearly proportional to the number of qubits in a chain following the previous work \cite{leung_robust_2018, murali2020architecting}. 
We use the latency of state detection \cite{burrell_scalable_2010} and electromagnetically induced transparency (EIT) cooling \cite{feng_efficient_2020} from the previous experimental results.
EIT cooling is used for sympathetic cooling, which reduces the phonon number of the qubit without disturbing the internal state.
For photonic entangling, we use the average latency of the heralding process (i.e., 1 MHz attempt rate with $2.5\times10^{-4}$ of success probability \cite{oreilly_fast_2024}) due to its probabilistic nature.
For shuttling, we set the shuttling latency as the sum of linear transportation, splitting, merging, and physical swapping based on the experimental results \cite{delaney_scalable_2024, pino_demonstration_2021, clark_characterization_2023}.
We calculate the time cost for the transportation to move the qubit for a unit distance (375 $\mu$m) with shuttling \cite{delaney_scalable_2024}.
We assume two types of transportation: fast and stable transportation.
Fast transportation can reduce the time cost with its high speed (82 $\mathrm{m/s}$ \cite{clark_characterization_2023}), but the qubit heating during the fast transportation can incur a high entangling-gate error.
Therefore, we use fast transportation only for the qubit movement for measurements, which is less susceptible to heating.
For other types of transportation (e.g., Bell state provisioning), we use stable transportation with 8~$\mathrm{m/s}$ speed \cite{delaney_scalable_2024}.

\subsubsection{Base error rate setup}{\label{subsubsection:OperationCost}}
\begin{table}
    \setlength{\tabcolsep}{6pt}
    \renewcommand{\arraystretch}{1.5}
    \footnotesize{
    \begin{tabular}{lcc}
    \hline\hline
    Operation Type & Error rate \\
    \hline
        Single-qubit gate      & $1.5 \times 10^{-7}$ \cite{smith_single-qubit_2024} \\
        Two-qubit gate         & $3.0 \times 10^{-4}$ \cite{loschnauer_scalable_2024} \\
        Measurement            & $9.0 \times 10^{-5}$ \cite{burrell_scalable_2010} \\
        Photonic entangling    & $2.85 \times 10^{-2}$ \cite{main_distributed_2024} \\
        Shuttling              & $1.0 \times 10^{-5}$ \cite{wright_reliable_2013} \\
        Idle error (for 1 $\mathrm{ms}$)   & $3.7 \times 10^{-6}$ \cite{reichardt_demonstration_2024} \\
    \hline\hline
    \end{tabular}}
    \caption{Base Error Rate.}
    \label{table:BaseErrorRate}
\end{table}

Table~\ref{table:BaseErrorRate} shows the error rate used in our evaluation.
All error rates are based on experimentally demonstrated results \cite{smith_single-qubit_2024,loschnauer_scalable_2024,burrell_scalable_2010,main_distributed_2024,wright_reliable_2013,reichardt_demonstration_2024} to evaluate the current reasonable quantum computers.
In this setup, we use the single-qubit gate and two-qubit gate error rates of $1.5\times10^{-7}$ and $3.0\times10^{-4}$, respectively, following the previous works \cite{smith_single-qubit_2024,loschnauer_scalable_2024}, while assuming that the platforms can support individual addressing \cite{chen_low-crosstalk_2024, hou_individually_2024, lim_design_2025, crain_individual_2014, shih_reprogrammable_2021, mehta_integrated_2020, mordini_multi-zone_2024, mehta_integrated_2016}.
For state detection error (\textit{i.e.}, measurement error), we refer to Ref.~\cite{burrell_scalable_2010}, which demonstrated high-fidelity measurement on the order of $10^{-5}$ error without relying on a heralded process~\cite{sotirova_high-fidelity_2024} or requiring excessively long measurement times~\cite{zhukas_high-fidelity_2021}.
For the photon entangling, shuttling, and coherence time, we use the results of the previous works demonstrating minimum errors \cite{main_distributed_2024,wright_reliable_2013,reichardt_demonstration_2024}.
Idle error arises from dephasing and technical imperfections affecting qubits in idle states, such as shuttling-induced phase noise and cross-talk. We adopt a dephasing-based error model, using parameters referenced from Ref.~\cite{reichardt_demonstration_2024}, to determine the idle error rate of $3.7 \times 10^{-6}$ per ms.

For future scenarios, we assess architecture-level performance gains by proportionally reducing the base error rates with the operation improvement factor $\lambda$.
We assumed uniform fractional improvements across all operations for simplicity.

\subsubsection{Logical error rate setup \label{subsubsec:logical_error_rate_setup}}

To estimate the logical error rates of logical two-qubit gates for each architecture, we perform circuit-level simulations of QEC processes, explicitly reflecting architecture-dependent operation sets.
Even though the same error correction code and decoding algorithm were used for all architectures, the actual operation sets for QEC circuits, including qubit shuttling and entanglement distribution, differ significantly, resulting in distinct error accumulation characteristics.
We implement architecture-specific QEC circuits using the Stim simulator~\cite{gidney_stim_2021} and apply the concatenated minimum-weight perfect matching (MWPM) decoder for syndrome decoding~\cite{lee_color_2025,lee_seokhyung-leecolor-code-stim_2025} (See Appendix~\ref{appendix:LogicalErrorRateCalculation}).

Each circuit consists of one transversal two-qubit gate followed by $d_\mathrm{code}$ rounds of syndrome extraction, corresponding to a single logical clock cycle.
The transversal gates are implemented by parallel remote two-qubit gates, and the error during the transversal gates propagates across different code blocks, which ideally requires sophisticated correlated decoders jointly involving multiple code blocks \cite{cain_correlated_2024}.
However, to simplify our analysis, we ignore such correlations and simulate error propagation within a single code block, where the ordinary concatenated MWPM decoder suffices.
Specifically, rather than including noisy two-qubit gates (with error rate $p$) between different code blocks explicitly, we place noisy idling gates with error rate $(4/5)p$ on all physical qubits in a code block.
This reflects the fact that 12 out of 15 nontrivial two-qubit Pauli operators acts nontrivially on each qubit.
We note that this simplification renders it difficult to evaluate logical error correlations between code blocks, which are therefore estimated conservatively in our application analysis (see Sec.~\ref{subsubsec:application_setup}).

The syndrome extraction circuit after the transversal gates is constructed in a way that heavily depends on the architecture and the code distance. 
Especially, for $d_\mathrm{code}>7$, DQC architectures requires partitioning of the logical qubit into multiple nodes, which significantly changes the actual operation sets of syndrome extraction.
The simulation circuits are constructed to reflect the architecture-dependent operation sets (SDQC, QCCD, and Photonic DQC) and are simulated up to code distance 13.

To evaluate logical error, the constructed circuit above is preceded and followed by logical initialization and measurement in the $Z$ basis for obtaining the logical $X$ error rate $p_\mathrm{L}^X$.
Then the total logical error rate $p_\mathrm{L}$ is calculated as $p_\mathrm{L} = 2 p_\mathrm{L}^X$, assuming that the logical $Z$ error rate is the same as $p_\mathrm{L}^X$ and logical $Y$ errors are negligible.

The transversal gates and syndrome extraction stages exhibit distinct physical characteristics and scaling behaviors in the studied architectures.
To characterize the effect of these two distinct stages, we introduce two parameters: the transversal gate error rate $p_{\mathrm{trans}}$ and the operation improvement factor for syndrome extraction $\lambda_\mathrm{SE}$.
The transversal gate error rate represents the error probability of two-qubit gates acting between code blocks, which are modeled as idling gates with error rates of $(4/5)p_\mathrm{trans}$, as described above.
For syndrome extraction, which consists of multiple operations with different physical error rates, we introduce a uniform reduction factor $\lambda_\mathrm{SE}$ that lowers all error rates rather than defining a single effective error rate.
For example, $\lambda_\mathrm{SE}=1$ corresponds to the current feasible error rates in Table~\ref{table:BaseErrorRate} during syndrome extraction, and $\lambda_\mathrm{SE}=10$ represents a tenfold improvement across all operations.
Although $p_{\mathrm{trans}}$ actually depends on the underlying gate error rates, we treat it as independent in our simulations and later assign appropriate values when evaluating application-level performance.

Assuming the effects of the two stages on logical error are independent, we model the logical error rate $p_{\mathrm{L}}$ as
\begin{align}
    p_{\mathrm{L}}(p_\mathrm{trans}, \lambda_\mathrm{SE}; d_\mathrm{code}) = A p_{\mathrm{trans}}^{\alpha d_\mathrm{code}} + B (1/\lambda_\mathrm{SE})^{\beta d_\mathrm{code}}, \label{equation:LogicalErrorRateModel}
\end{align}
where $d_\mathrm{code}$ is the code distance and $A$, $B$, $\alpha$, and $\beta$ are fitting parameters.
Note that we fit data for each fixed $d_\mathrm{code}$ with varying $p_\mathrm{trans}$ and $\lambda_\mathrm{SE}$, indicating that the parameters are functions on $d_\mathrm{code}$ (although not explicitly specified for brevity).
The model suggests two regimes: the transversal gate dominated regime and syndrome extraction dominated regime.
For large $p_{\mathrm{trans}}$, the logical error rate is primarily determined by transversal gate errors and increases with $p_{\mathrm{trans}}$.
On the other hand, when $p_{\mathrm{trans}}$ becomes sufficiently small, the logical error rate is saturated and limited by the fidelity of syndrome extraction stage.
The crossover error rate $p^*_{\mathrm{trans}} = (B/A\lambda_\mathrm{SE}^{\beta d_\mathrm{code}})^{1/(\alpha d_\mathrm{code})}$, defined as the point where the two contributions become equal, marks the boundary between the two regimes and can be obtained by fitting the model to simulation results.

Due to the limited computational resources, our simulation was performed down to the level of $10^{-8}$ logical error rate, and the lower regions were extrapolated from the fit parameters.
The results are presented in Appendix~\ref{appendix:QECSimulationResult} together with the values of the fitted parameters, the coefficient of determination ($R^2$), and $p_{\mathrm{trans}}$ at $\lambda_\mathrm{SE}=1$, verifying that the model represents the data reasonably well within our simulated range.
Although the fits become less accurate for Photonic DQC at $d_\mathrm{code} \ge 7$ due to noisy photonic entangling operations, they still yield slightly optimistic estimates relative to the raw data, suggesting that the comparison between SDQC and Photonic DQC remains conservative.

\subsubsection{Application setup \label{subsubsec:application_setup}}

\begin{table*}
  \setlength{\tabcolsep}{6pt}
  \renewcommand{\arraystretch}{1.5}
  \begin{tabular}{lccccc}
  \hline\hline\\[-1em]
  Application&\parbox{3cm}{Number of\\logical qubits}&\parbox{2.2cm}{Number of logical layers per shot ($N_{\mathrm{layer}}$)}&\parbox{2cm}{Number of logical gates per shot ($N_{\mathrm{gate}}$)}&\parbox{2.5cm}{Number of idle operations \\per shot \\($N_{\mathrm{idle}}$)}&\parbox{2cm}{Number of shots \\($N_{\mathrm{shots}}$)}\\[1em]
  \hline
  Fermi-Hubbard \cite{jafarizadeh_recipe_2024}&$132$&\num{8787}&\num{331024}&\num{828860}&\num{10000} shots\\
  ECDLP \cite{haner_improved_2020}&\num{2871}&$1.4\times2^{27}$&$1.4\times2^{34}$&$4.91\times10^{11}$&\num{1} shot\\
  \hline\hline
  \end{tabular}
  \caption{Application setup.}
  \label{table:ApplicationSetup}
\end{table*}

Overall architectural performance can be evaluated by the estimated execution time and the algorithmic success probability.
We benchmark two representative applications: quantum simulation of the Fermi-Hubbard model with Trotterization \cite{lloyd_universal_1996, jafarizadeh_recipe_2024} and the elliptic-curve discrete logarithm problem (ECDLP) implemented with Shor's algorithm \cite{haner_improved_2020}.
For the Fermi-Hubbard model, we simulate a $6\times8$ lattice of spinful fermions, requiring 132 logical qubits \cite{jafarizadeh_recipe_2024}.
Each shot comprises \num{8787} layers of logical two-qubit gates (\num{14548} gates) and non-Clifford gates to compose arbitrary rotation gates (\num{316476} $T$ gates for $10^{-5}$ precision error \cite{kliuchnikov2023shorter}), and $10^4$ shots are assumed to suppress statistical error.
For ECDLP, we target a 256-bit elliptic-curve cryptography (ECC) instance, requiring \num{2871} logical qubits \cite{haner_improved_2020}.
The circuit depth is approximately $1.88\times10^8$ logical layers, requiring approximately $2.41\times10^{10}$ logical gates, including single-qubit non-Clifford gates.
For each architectures, the number of idle operations is estimated by summing idle logical qubits across logical layers.
Table~\ref{table:ApplicationSetup} summarizes the application setup.

\begin{align}
T_{\mathrm{exec}} &= N_{\mathrm{layer}} \times T_{\mathrm{L}} \times N_{\mathrm{shots}}, \label{eq:ExecTime}
\end{align}
\begin{align}
P_{\mathrm{success}} &=
    (1 - 2p_{\mathrm{L}})^{N_{\mathrm{gate}}} \times (1 - p_{\mathrm{idle}})^{N_{\mathrm{idle}}}.\label{eq:SuccessRate}
\end{align}

Equations \eqref{eq:ExecTime} and \eqref{eq:SuccessRate} show the expressions for the total execution time $T_\text{exec}$ and the overall success rate $P_\text{success}$ of the benchmark applications.
The execution time $T_\text{exec}$ is obtained by multiplying the number of gate layers for each shot $N_{\mathrm{layer}}$, the number of shots for executing the target application $N_{\mathrm{shots}}$, and the time duration required for executing a logical operation $T_{\mathrm{L}}$ for each architecture.
The success rate $P_\text{success}$ accounts for both the logical operation error rate $p_{\mathrm{L}}$ and the idle operation error rate $p_{\mathrm{idle}}$.
We calculate $p_{\mathrm{idle}}$ as the logical qubit error without transversal gate error rate (i.e., $p_{\mathrm{L}}(0, \lambda_\mathrm{SE}; d)$) because the transversal operations are not applied during an idle operation.
The total success rate is computed by accumulating the success probabilities of all operations, weighted by their counts (i.e., total number of logical gates $N_{\mathrm{gate}}$, and idle operations $N_{\mathrm{idle}}$).
We conservatively increase the logical operation error by twice because a single logical operation can incur the logical error of up to two logical qubits.
Note that both $T_{\mathrm{L}}$ and $p_{\mathrm{L}}$ depend on the architecture, the number of logical qubits, and the code distance.
A target success rate of 90~\% is set for both applications, and the required code distances and gate error improvements are analyzed for each architecture.

\subsection{Operation time cost}{\label{subsection:TimeCost}}

\begin{figure*}
  \includegraphics[width=0.8\paperwidth]{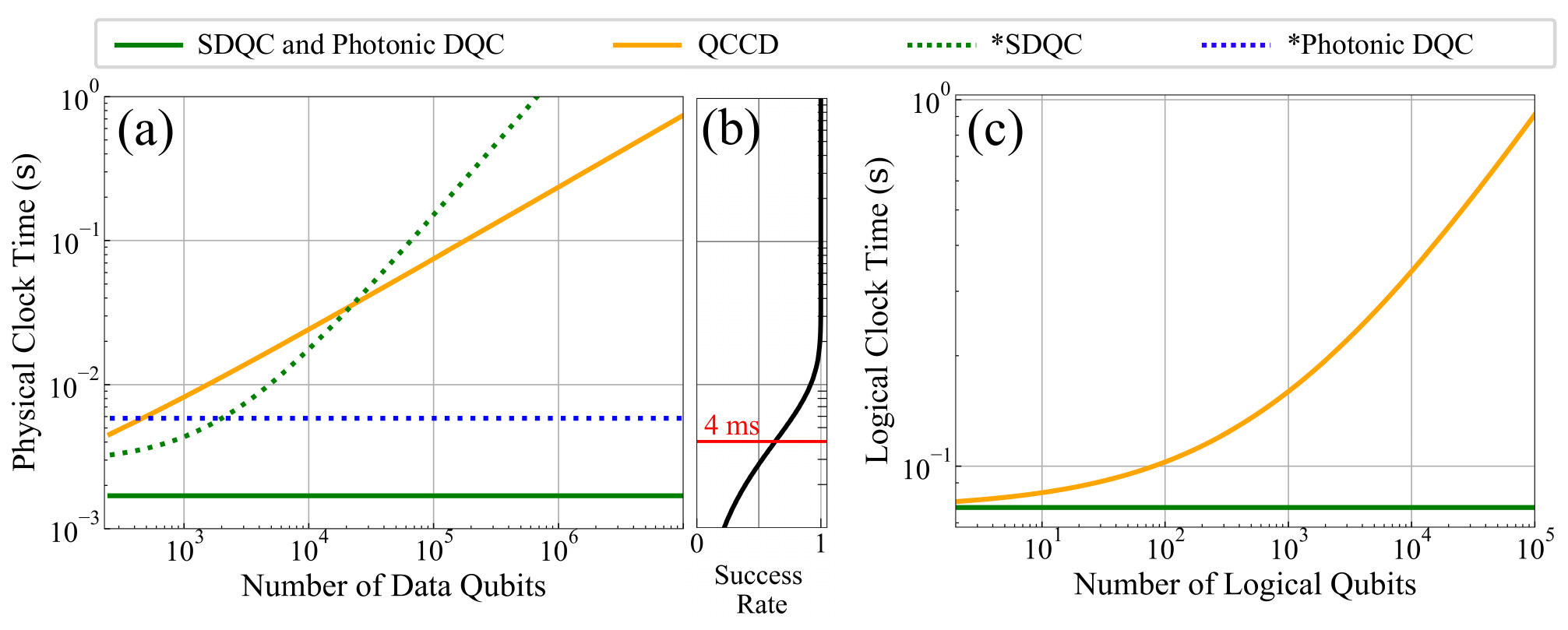}
  \caption{Time cost with $d_{code}=13$. 
  (a) The remote physical two-qubit gate time (physical clock time) versus the number of data qubits.
  The dotted lines (*SDQC and *Photonic DQC) represent the time costs including entanglement distribution time.
  The green solid line shows the identical time cost for SDQC and Photonic DQC when the entanglement distribution time is removed by pipelining, exhibiting scale-independent behavior.
  The orange solid line shows the physical two-qubit gate time of QCCD, in which pipelining is not applicable.
  (b) The cumulative success probability of photonic entangling as a function of attempt duration.
  Photonic entangling is a probabilistic process following a geometric distribution, and the red line indicates the mean value ($4~ms$) used for architecture evaluation.
  (c) The logical clock time versus the number of logical qubits.
  For all architectures, the syndrome extraction qubits distribution is assumed to be fully pipelined.
}{\label{fig:TimeCost_d13}}
\end{figure*}

In this section, we compare three architectures with the time cost of remote physical two-qubit gates (inter-node two-qubit gates) and logical gates with QEC cycles.

\subsubsection{Remote physical two-qubit gates}{\label{subsubsection:TimeRemoteTQ}}
Figure~\ref{fig:TimeCost_d13}(a) presents the execution time analysis of remote physical two-qubit gates across three architectures.
As the number of qubits increases, the total shuttling distance also increases in the two shuttling-based architectures, SDQC without pipelining (denoted as *SDQC in Fig.~\ref{fig:TimeCost_d13}) and QCCD, resulting in longer execution times for remote physical gates.
In QCCD, the shuttling distance scales with the square root of the total number of qubits due to its two-dimensional grid structure.
This distance can be mitigated through routing optimization, and our evaluation adopts the strategy proposed in Ref. \cite{webber_efficient_2020}.
In contrast, for *SDQC, the time cost scales linearly with system size (\textit{i.e.}, the number of logical qubits), as it corresponds to a combinational average over all logical qubits.
Details are discussed in Appendix \ref{appendix:TimeCostCalculation}.

Crucially, QCCD requires the shuttling of data qubits, which cannot be pipelined and therefore contributes directly to cumulative latency.
In *SDQC, however, the increase in shuttling time arises solely from entanglement distribution.
This overhead does not accumulate over successive two-qubit gates with a proper pipelining strategy, denoted as SDQC in Fig.~\ref{fig:TimeCost_d13} (see Section \ref{subsection:Piplining}).
Photonic DQC also utilizes pre-distributed entangled qubits for remote physical two-qubit gate, and such operations do not scale with the system size since the photonic qubits propagate at the speed of light; *Photonic DQC in Fig.~\ref{fig:TimeCost_d13}(a) does not consider pipelining.
With pre-distributed entangled qubits via proper pipelining, both SDQC and Photonic DQC exhibit scale-independent execution times for remote physical two-qubit gates.
These execution times are effectively identical, as the two architectures share the same operational structure except for the entanglement distribution mechanism.

While both SDQC and Photonic DQC utilize distributed entanglement pairs to enable gate teleportation, their entanglement distribution mechanisms differ fundamentally.
SDQC employs deterministic shuttling of entangled gate teleportation qubits between nodes, enabling reliable delivery on demand.
In contrast, Photonic DQC relies on a heralded entanglement generation process, which involves probabilistic photon-mediated attempts repeated until success.
As a result, the entanglement generation time in Photonic DQC can only be characterized statistically, and its probability distribution depends on the success rate of individual entanglement generation attempts (Figure~\ref{fig:TimeCost_d13}(b)) \cite{oreilly_fast_2024}.
This probabilistic nature of Photonic DQC may necessitate additional buffering and resource scheduling, potentially resulting in increased execution time.

\subsubsection{Logical gates with QEC cycles}{\label{subsubsection:TimeLogicalTQ}}
For fault-tolerant gate operations, each logical clock cycle involves transversal gates followed by $d_\text{code}$ rounds of syndrome extraction for error-correction. In the case of DQC architectures, we adopt the pipelined execution strategy discussed in Section \ref{subsection:Piplining} and Appendix \ref{appendix:Pipelining}.
In all architectures, transversal gates can be performed in a fully parallel manner, and their execution time is essentially identical to the remote physical gate time discussed above.
The syndrome extraction is mostly scale-independent, as error correction codes are encoded into localized clusters of physical qubits.

Figure~\ref{fig:TimeCost_d13}(c) shows the logical clock time across the three architectures.
Similar to the behavior of physical two-qubit gates, both DQC architectures exhibit no scale-dependence.
In contrast, the logical clock cycle of QCCD increases with the number of logical qubits, due to longer shuttling paths.
As a result, DQCs exhibit the fastest logical clock speed, outperforming QCCD due to their ability to pipeline entanglement distribution, thereby reducing latency in transversal gates execution.
For $2,871$ logical qubits to run Shor's algorithm \cite{haner_improved_2020}, as follows Section \ref{subsection:Applications}, the logical clock speed of DQCs is 2.82 times faster than QCCD.

Although pipelining enables DQC architectures to achieve scale-independent logical clock speeds, this advantage holds only if entangled pair generation throughput exceeds the demand during logical operations.
Here, the entanglement pair throughput is defined as the number of entangled pairs generated per unit time at a single entanglement factory, while the peak entanglement demand refers to the highest entanglement generation rate required by any single factory.
If the available throughput falls short of this peak demand, pipelining fails to fully mask the latency of entanglement distribution, resulting in execution bottlenecks.

In SDQC, entangled pairs are generated using local two-qubit gates within each factory and distributed via shuttling, both of which can operate in parallel.
For the code distance $d_\textrm{code} = 13$, up to 198 pairs are required in each factory per FT cycle, resulting in a peak entanglement demand of \num{2568} Hz. 
The achievable throughput with a chain capacity of 60 is estimated at \num{39958} Hz, which exceeds this demand by more than an order of magnitude.

In Photonic DQC, the maximum number of required entangled pairs is slightly higher (total 276 pairs with locally generated 91 pairs) due to additional gate teleportation for Bell measurements of segmented syndrome extraction qubits, resulting in a peak entanglement demand of \num{3580} Hz; for only photonic entangling, the peak entanglement demand is \num{2399} Hz.
Given the current record generation rate of 250 Hz, achieving pipelined execution requires substantial improvements in both the speed and parallelization of entanglement generation.

\subsection{Error rate}{\label{subsection:ErrorRate}}
In this section, we evaluate the remote physical two-qubit gate error and logical two-qubit gate error of each architecture scaled by the number of qubits.

\subsubsection{Remote physical two-qubit gates}{\label{subsubsection:PhysicalError}}

\begin{figure*}
  \includegraphics[width=0.8\paperwidth]{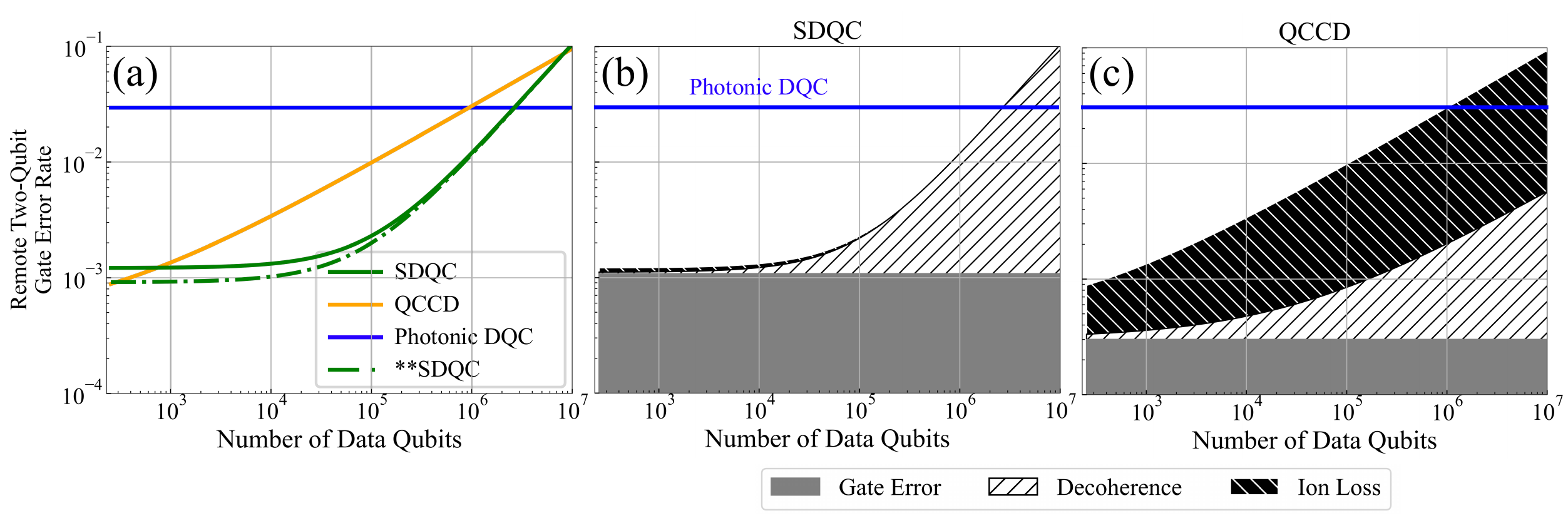}
  \caption{Remote physical two-qubit gate error rate.
  Remote physical two-qubit gate error rates used for transversal gates in each architecture with $d_\mathrm{code}=13$ and their error budgets.
  (a) Physical two-qubit gate error rates scaled by the number of physical qubits.
  **SDQC (the dashed-dotted line) represents the case in which the entanglement pair preparation error has been eliminated via entanglement purification, corresponding to the optimistic scenario.
  (b, c) Error budgets of SDQC (b) and QCCD (c). The error is analyzed into three categories: gate operation error, ion loss, and decoherence.
  The blue line indicates the remote two-qubit gate error rate of Photonic DQC, where photonic entangling error is the dominant contribution.
  }{\label{fig:ErrorRate_d13}}
\end{figure*}

Figure~\ref{fig:ErrorRate_d13} shows the average remote physical two-qubit gate error for each architecture, with the error budget decomposed into its constituent sources.
In all architectures, the total gate error is determined by three distinct error mechanisms: (1) accumulated errors from local gates and state detection, (2) ion loss at junctions during shuttling, and (3) decoherence over the total operation time.
Because the shuttling path length and operation time vary depending on the architecture and system scale, we constructed a simplified error model that captures the scale-dependent gate performance of each architecture for comparison (See Appendix \ref{appendix:PhysicalErrorRate}).

In the two DQC architectures, the gate teleportation protocol introduces additional local gate operations, increasing the baseline error rate.
In particular, the error of photonic entangling is approximately two orders of magnitude larger than that of local two-qubit gates, resulting in significantly higher error rates in Photonic DQC architecture.
In contrast, QCCD architecture requires only a single two-qubit gate after reconfiguring connectivity through qubit shuttling, yielding a lower gate error at a small scale.

As the number of qubits increases, the number of junctions along the shuttling path increases in the two shuttling-based architectures, which increases the operation error due to ion loss.
In QCCD, these quantities scale approximately with the square root of the number of qubits \cite{webber_efficient_2020}, and the growing number of junctions becomes a dominant source of physical gate error as the system scales.
In contrast, in SDQC, data qubits remain stationary and are thus immune to ion loss during shuttling.
gate teleportation qubits used for remote operations do not carry any quantum circuit information prior to their entangling gate with the data qubits.
As a result, they can be replaced with spare ones when ion loss occurs, significantly suppressing the chance of operational failure by ion loss, which scales exponentially with the number of spare qubits (see Appendix \ref{appendix:SpaceCostCalculation}).
As a result, the error due to ion loss during entanglement distribution in SDQC becomes negligible, in contrast to QCCD.
In addition, if sufficient time slack is available, the entanglement purification can be used to remove errors of entanglement generation, as indicated by the dashed-dotted line in Fig.~\ref{fig:ErrorRate_d13} (**SDQC).

The operational time also increases as the qubit number grows (see Section \ref{subsection:TimeCost}), which introduces decoherence-induced error, particularly in large-scale systems.
When evaluating decoherence-induced error, the entanglement distribution time, though hidden in the time cost by pipelining, still contributes, because the qubit coherence decays during this interval.
At extremely large scales (on the order of $10^6$ data qubits), the scale-dependent gate error of shuttling-based architectures becomes comparable to that of Photonic DQC.

The three architectures exhibit distinct scaling behaviors in their remote two-qubit gate errors.
SDQC depends only on decoherence, as gate teleporation qubits can be replaced when ion loss occurs.
Because the decoherence contribution remains smaller than the operational error over a wide range of system sizes, scale-dependent growth becomes apparent only at very large scales.
In contrast, QCCD experiences error growth across all scales due to shuttling-induced ion loss, whose error rate is comparable to operational error even at small system size. 
Decoherence further contributes to the overall error at large scales, though its impact remains secondary to shuttling-induced loss.
Photonic DQC shows no scale dependence, and hence its error rate remains constant with system size.
However, the intrinsic photonic entangling error is substantially higher than the local operational achievable in trapped-ion systems.

In summary, although QCCD architecture achieves low gate error at small system sizes, as the system scales, SDQC outperforms it in the intermediate-to-large scale regime ($1000 < n_{\text{ph}} < 10^7$).
Furthermore, the gate error of SDQC can be further suppressed by up to a factor of $1.33$ through the use of entanglement purification during the pipelined distribution process.

\subsubsection{Logical Gates with QEC cycles}{\label{subsubsection:LogicalError}}

\begin{figure*}
  \includegraphics[width=0.7\paperwidth]{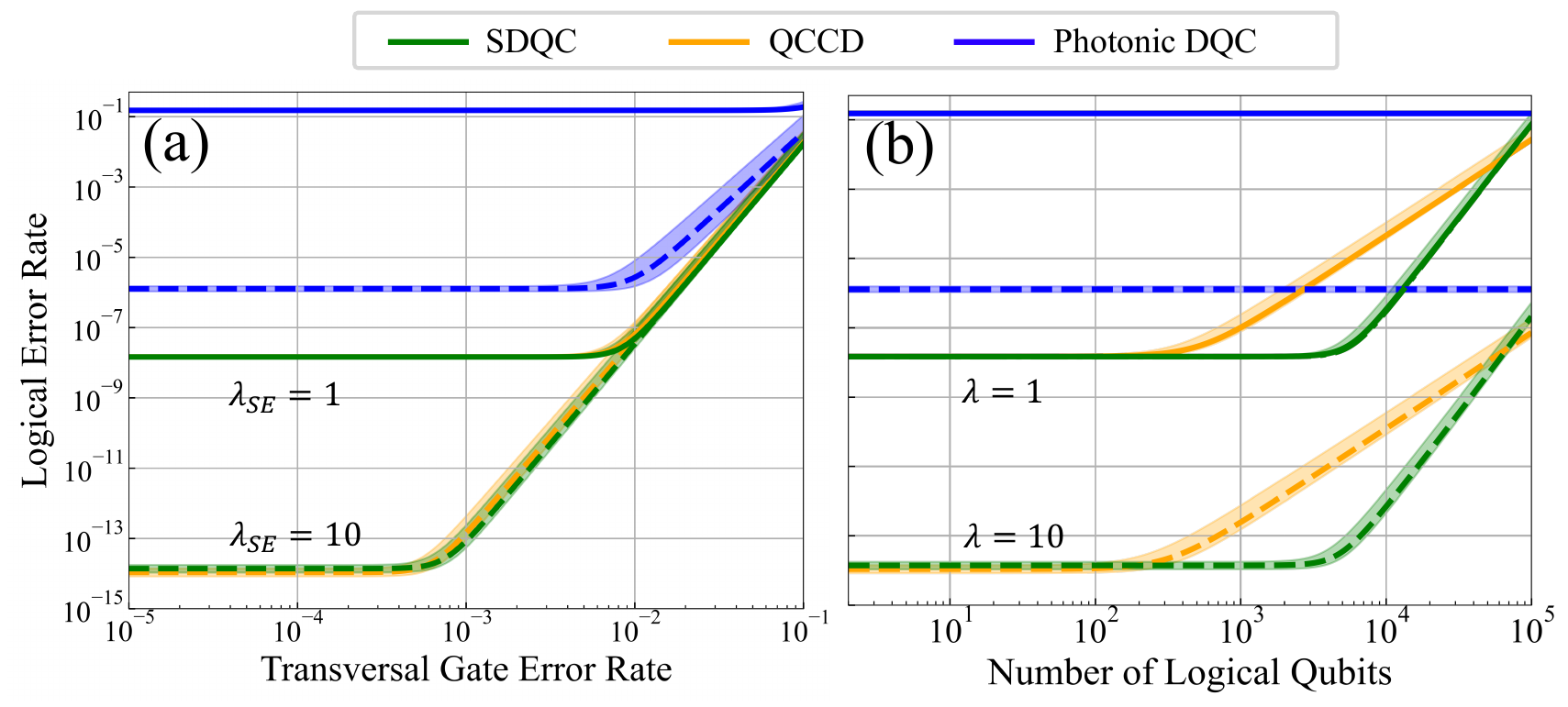}
    \caption{
        Logical two-qubut gate error rate.
        Logical error rates $p_\mathrm{L}$ of transversal two-qubit gates followed by $d_\mathrm{code}$ rounds of syndrome extraction for code distance $d_\mathrm{code}=13$ across all architectures, computed based on the setup shown in Sec.~\ref{subsubsec:logical_error_rate_setup}.
        (a) Logical error rates as a function of the transversal gate error rate ($p_\mathrm{trans}$) for different operation improvement factors for syndrome extraction ($\lambda_\mathrm{SE}=1,10$), where the fitting parameters in Table~\ref{table:LogicalErrorRateFittingParameters} are used for extrapolation.
        The color bands represent the upper/lower bounds with the standard deviations of each parameter.
        (b) Logical error rates of each logical qubit as a function of the number of logical qubits, for different operation improvement factors $\lambda=1,10$. Here, the number of logical qubits determines $p_\mathrm{trans}$ as described in Section~\ref{subsubsection:PhysicalError}.
    }
    \label{fig:LogicalErrorRate}
\end{figure*}

A logical two-qubit gate is implemented based on transversal gates of parallel remote two-qubit gates, followed by QEC cycles.
Based on the results of the previous section and QEC simulations, we estimate how the logical error rate scales with system size for each architecture.

Figure~\ref{fig:LogicalErrorRate}(a) shows the obtained logical error rates as a function of the transversal gate error rate $p_\mathrm{trans}$, for code distance $d_\mathrm{code}=13$ across all architectures.
We analyze logical error rates with two operation improvement factors for syndrome extraction, $\lambda_\mathrm{SE} = 1$ and $\lambda_\mathrm{SE} = 10$, representing the current performance and a tenfold enhancement in operation fidelity during syndrome extraction.
Photonic DQC exhibits a broad plateau at a relatively high logical error rate, indicating that its performance is limited by errors in syndrome extraction.
This limitation arises because segmented stabilizers for Photonic DQC are measured using entangled qubits mediated by photonic interconnections, whose relatively low entanglement fidelity leads to a high logical error rate of $6.36^{+1.19}_{-1.02}\times10^{-7}$ for the $\lambda_\mathrm{SE}=10$ case.
In contrast, both SDQC and QCCD achieve significantly lower logical error rates in the syndrome extraction dominated regime, yielding $6.89^{+1.93}_{-1.55}\times10^{-15}$ and $5.41^{1.62}_{-1.28}\times10^{-15}$ at $\lambda_\mathrm{SE}=10$, respectively.
The corresponding crossover points $p^*_{\text{trans}}$ for SDQC are $6.46^{+0.91}_{-1.17}\times10^{-3}$ and $5.63^{+1.21}_{-1.72}\times10^{-4}$ for $\lambda_\mathrm{SE} = 1$ and $10$, respectively, whereas the currently available two-qubit gate fidelities lie below these crossover points..
This implies that further improving the transversal gate error rate below $p^*_{\text{trans}}$ would not yield additional gains in logical performance.

From the simulation results, we estimate how the logical error rate scales with the system size for each architecture, as shown in Fig.~\ref{fig:LogicalErrorRate}(b).
Here, the transversal gate error rate is taken from the remote two-qubit gate error analysis in Section~\ref{subsubsection:PhysicalError}.
Since the syndrome extraction error depends only on the code distance and architecture type, it remains independent of the number of logical qubits.
Unlike in Fig.~\ref{fig:LogicalErrorRate}(a), we apply the same operation improvement factor $\lambda$ to both the transversal gate and syndrome extraction errors to assess the overall effect of enhanced qubit-control fidelity.
At small system sizes, SDQC exhibits a slightly higher remote two-qubit gate error rate than QCCD, yet both achieve nearly identical logical error rates because the performance is dominated by syndrome extraction errors.
As the system scales up, the remote gate error in QCCD increases more rapidly than in SDQC, causing QCCD to enter the transversal-gate-dominated regime earlier.
Although SDQC eventually crosses the same crossover point, it maintains lower logical error rates over a wide range of system sizes, from $10^2$ to $10^5$ logical qubits, sufficient for most practical applications.
Notably, the range in which SDQC outperforms QCCD is almost independent of the overall improvement factor, indicating that SDQC consistently yields better fault-tolerance scalability.
The logical error rate of Photonic DQC is scale-independent, as is the physical two-qubit gate error rate, and lies within the syndrome-extraction-dominated regime.

\subsection{Applications}{\label{subsection:Applications}}

\begin{figure*}
  \includegraphics[width=0.7\paperwidth]{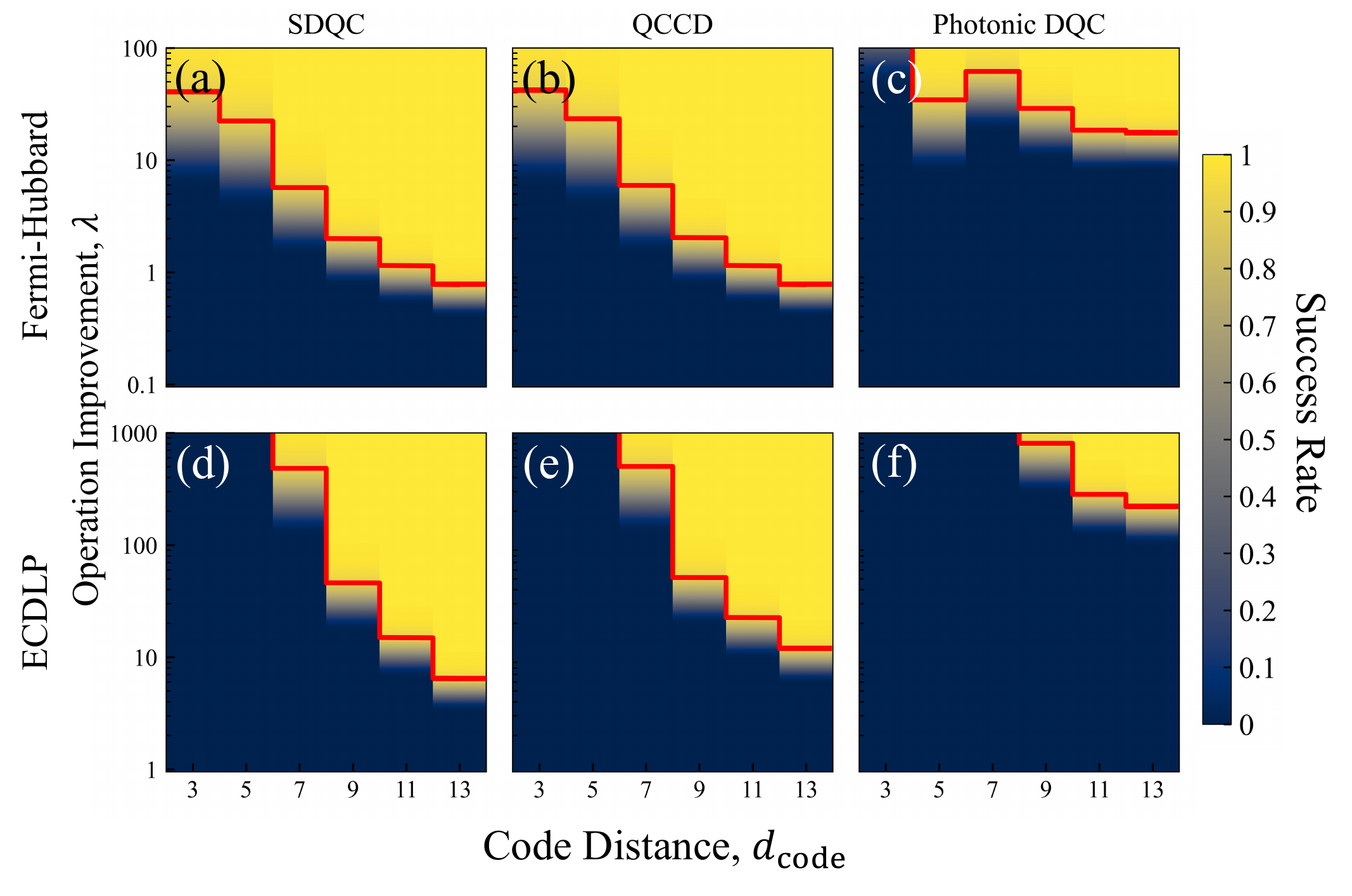}
  \caption{Application success rate as a function of code distance and operation improvement factor $\lambda$. 
  Panels (a)-(c) show the success rates of SDQC, QCCD, and Photonic DQC for the Fermi-Hubbard model, and panels (d)-(f) show the corresponding results for ECDLP.
  The red line denotes the boundary for achieving a 90~\% success rate. }{\label{fig:ApplicationSuccessRate}}
\end{figure*}

\begin{figure*}
  \includegraphics[width=0.7\paperwidth]{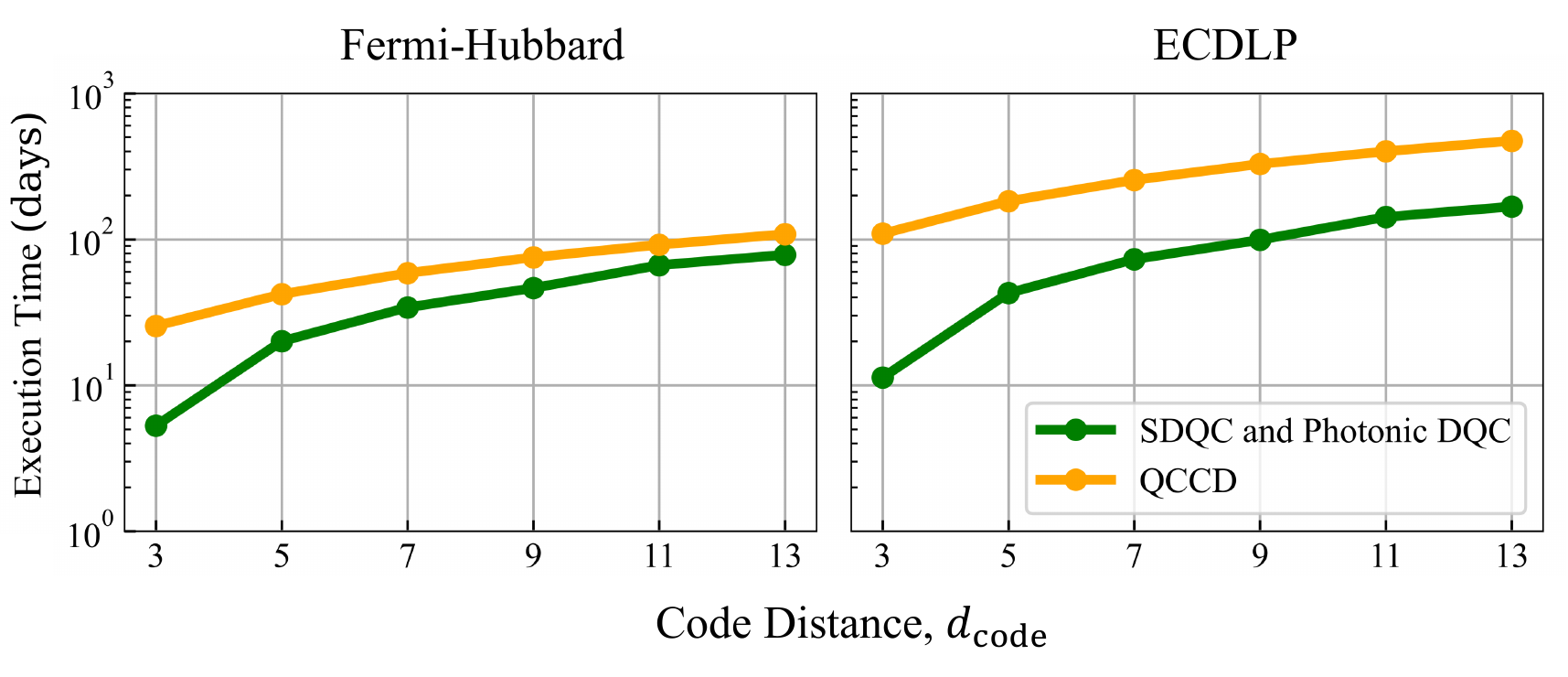}
  \caption{Application execution time as a function of code distance. Panels (a) and (b) show the execution time of three architectures for the Fermi-Hubbard model simulation and ECDLP.}
  {\label{fig:ApplicationCost}}
\end{figure*}

\begin{table*}
  \setlength{\tabcolsep}{6pt}
  \renewcommand{\arraystretch}{1.5}
  \begin{tabular}{lccccccc}
  \hline\hline\\[-1.5em]
  \multirow{2}{*}{Application}&\multirow{2}{*}{Architecture}&\multicolumn{4}{c}{Space Cost}&\multirow{2}{*}{\parbox{1.5cm}{Success\\Rate\\(\%)}}&\multirow{2}{*}{\parbox{1.5cm}{Execution\\Time (days)}}\\[-0.2em]
  &&Data&\parbox{1.5cm}{Syndrome\\Extraction}&\parbox{2cm}{Gate\\Teleportation}&Total&&\\[0.5em]
  \hline
  \multirow{3}{*}{\parbox{1.2cm}{Fermi-Hubbard \cite{jafarizadeh_recipe_2024}}}&SDQC&\num{16764}&\num{33264}&\num{11424}&\num{61452}&$98.91^{+0.10}_{-0.10}$&78\\
  &QCCD&\num{16764}&\num{33264}&\num{0}&\num{50028}&$98.90^{+0.10}_{-0.12}$&108\\
  &Photonic DQC&\num{16764}&\num{52800}&\num{10668}&\num{80232}&$0$&78\\
  \hline
  \multirow{3}{*}{\parbox{1.2cm}{ECDLP \cite{haner_improved_2020}}}&SDQC&\num{364617}&\num{723492}&\num{96040}&\num{1184149}&$99.62^{+0.08}_{-0.48}$&168\\
  &QCCD&\num{364617}&\num{723492}&\num{0}&\num{1088109}&$90.44^{+3.32}_{-25.85}$&473\\
  &Photonic DQC&\num{364617}&\num{1148400}&\num{87122}&\num{1600139}&$0$&168\\
  \hline\hline
  \end{tabular}
  \caption{Space cost, success rate, and execution time of each architecture to run Fermi-Hubbard model and ECDLP in $d_\mathrm{code}=13$. We evaluate the architectures under $\lambda=1$ and  $\lambda=10$ for Fermi-Hubbard and ECDLP, respectively, where at least one of the architectures can successfully achieve higher than our target success rate (90~\%).
}
  \label{table:ApplicationResults}
\end{table*}

In this section, we evaluate architectures in terms of success rate, execution time, and space cost when executing applications.
We evaluate three architectures in various operation improvement factors ($\lambda$; from 0.1 to 1000 times of improvements) and code distances ($d_\mathrm{code}=\{3,5,7,9,11,13\}$).
The success rate and execution time are calculated by using Eq.~\eqref{eq:ExecTime} and Eq.~\eqref{eq:SuccessRate}, respectively.
The space cost is calculated as the total number of physical qubits including data qubit count, qubit count for syndrome extractions, and qubit count for gate teleportation.
We evaluate the space cost under $\lambda=1$ and  $\lambda=10$ for Fermi-Hubbard and ECDLP, respectively, where at least one of the architectures can successfully achieve higher than our target success rate (90~\%) for each application.

First, Figure~\ref{fig:ApplicationSuccessRate} shows the success rates of the two applications (i.e., Fermi-Hubbard model and ECDLP) in various operation improvement factors $\lambda$ and code distances $d_\mathrm{code}$.
For the Fermi-Hubbard model, SDQC and QCCD do not require substantial operation improvement factors and large code distances due to the relatively small scale of the application.
As a result, SDQC and QCCD can execute the Fermi-Hubbard model with $98.91^{+0.10}_{-0.10}~\%$ and $98.90^{+0.10}_{-0.12}~\%$ of success rates, under $d_\mathrm{code}=13$ and $\lambda=1$.
In contrast, Photonic DQC requires a significant improvement factor (i.e., $\lambda>10$) to successfully run Fermi-Hubbard model. 
In addition, Photonic DQC does not show a clear trend of gate-improvement reduction to achieve the target success rate beyond $d_\mathrm{code}=5$.
These results are originated from the erroneous photonic entanglements of Photonic DQC.
The results also highlight the need to improve photonic entangling fidelity and the importance of reducing the use of photonic entanglements in architectural designs.

On the other hand, for ECDLP, both SDQC and QCCD require a higher operation improvement factor, due to the higher number of logical gates of ECDLP than Fermi-Hubbard (as shown in Table~\ref{table:ApplicationSetup}).
For instance, SDQC and QCCD can execute the ECDLP with $99.62^{+0.08}_{-0.48}~\%$ and $90.44^{+3.32}_{-25.85}~\%$ of success rates with $d_\mathrm{code}=13$, respectively, under $\lambda=10$.
In the case of Photonic DQC, it requires an even larger improvement factor ($\lambda=200$) and code distance ($d_\mathrm{code}=13$) to successfully run ECDLP due to its erroneous photonic entanglement.
Overall, the success rate results highlight the advantage of SDQC over other architectures because SDQC requires lower error improvements and code distances to achieve the target success rate (90~\%).

Second, Figure~\ref{fig:ApplicationCost} shows the execution time of three architectures for various code distances ($d_\mathrm{code}=\{3,5,7,9,11,13\}$).
The execution time of SDQC and Photonic DQC is identical and shorter than that of QCCD.
The shorter execution time of both DQCs originates from the use of pipelining in entanglement distribution, which makes SDQC and Photonic DQC hide the long latency of entanglement distribution.
As a result, compared to QCCD, SDQC achieves \num{4.82} times (at $d_\mathrm{code}=3$) to \num{1.38} times (at $d_\mathrm{code}=13$) shorter execution time in Fermi-Hubbard model, and \num{9.66} times (at $d_\mathrm{code}=3$) to \num{2.82} times (at $d_\mathrm{code}=13$) shorter execution time in ECDLP.
As the logical clock time of QCCD increases with the number of logical qubits used in applications (as shown in Fig.~\ref{fig:TimeCost_d13}), SDQC achieves a higher speed-up at ECDLP, requiring a larger number of qubits than Fermi-Hubbard model.
Note that, although Photonic DQC shows the same execution time as SDQC, Photonic DQC suffers from a much lower success rate than SDQC for the same error improvement and code distance, as shown in Fig.~\ref{fig:ApplicationSuccessRate}.

Finally, Table~\ref{table:ApplicationResults} shows the space cost, success rate, and execution time of three architectures in $d_\mathrm{code}=13$ for Fermi-Hubbard model and ECDLP, respectively.
We evaluate the architectures under $\lambda=1$ for Fermi-Hubbard model and  $\lambda=10$ for ECDLP, where at least one of the architectures achieves the target success rate for the first time.
Overall, SDQC shows comparable space cost to the other architectures while achieving the highest success rate and lowest execution time.
For Fermi-Hubbard model, SDQC shows only 22.84~\% higher space cost than QCCD, thanks to the relatively small portion of gate teleportation qubits in the total qubit count.
In the case of ECDLP, SDQC shows 8.83~\% higher space cost than QCCD in $d_\mathrm{code}=13$, while achieving 64.48~\% lower execution time and $9.18^{+25.93}_{-3.80}~\%$ higher success rate than QCCD.
Note that the execution time gap between SDQC and QCCD widens for applications requiring a larger system scale (i.e., ECDLP).
These results demonstrate the significant potential of SDQC for large-scale applications in the FTQC era.

\section{Conclusion}{\label{Section:Conclusion}}
In this work, we proposed a scalable shuttling-based distributed quantum computing architecture, SDQC, for trapped-ion quantum computers.
SDQC leverages the key advantages of both Photonic DQC and QCCD; it utilizes asynchronous entanglement distribution to achieve scale-independent time cost, while employing deterministic shuttling to deliver entangled qubit pairs with high fidelity. 
We designed a QEC scheme with segmented stabilizers, which enables encoding a logical qubit into multiple processor nodes without local connectivity, and developed a pipelined FT-cycle execution enabling high logical clock speed.
As a result, SDQC exhibits fast and scale-independent logical clock speed and a high success rate of running practical quantum algorithms with modest space cost overhead, outperforming other scalable trapped-ion architectures.
Future work may include extending SDQC to support fault-tolerant non-Clifford operations by architectural design of magic state preparation, and investigating the architecture performance across various QEC codes to explore code–architecture co-optimization opportunities.

\begin{acknowledgments}
The authors thank Kenneth R. Brown and Mingyu Kang for insightful discussions.
Technical support from Hosung Seo is gratefully acknowledged.
The research contributions introduced in this article was supported in part by National Research Foundation of Korea(NRF) (Grant No. RS-2025-16065694, RS-2023-NR068116, RS-2022-NR068814, RS-2023-00302576, RS-2024-00466865, 2022M3E4A105136512) and Information \& Communications Technology Planning \& Evaluation(IITP) (Grant No. RS-2025-25442569, RS-2022-II221040) funded by the Korea government(MSIT).
SHL is supported by the Australian Research Council via the Centre of Excellence in Engineered Quantum Systems (EQUS) Project No. CE170100009.
\end{acknowledgments}

\appendix
\begin{widetext}
\section{Architectural assumptions and feasibility of SDQC}
\label{appendix:assumption_sdqc}

This study is based on several assumptions for SDQC, which are detailed below. These assumptions are validated through component-wise demonstrations, as summarized in Table~\ref{table:ComponentWiseDemonstration}, thereby supporting their feasibility.

\begin{enumerate}
    
  \item SDQC can apply gate operations to qubits inside processor nodes, entanglement factories, and detectors.
In addition, SDQC supports individual addressing for gate operations.

  \item SDQC can apply gate operations in parallel to the qubits within the chain.
SDQC can drive the single-qubit gates to qubits in parallel because each ion does not share its internal states with other qubits.
SDQC also can drive multiple two-qubit gates in parallel by modulating the driving fields. This technique ensures that only the target ion pairs are entangled at the end of the operation \cite{figgatt_parallel_2019, grzesiak_efficient_2020, chen_benchmarking_2024}.

  \item SDQC can shuttle many communication ions in parallel without deadlock.
Multiple potential wells, driven by dynamic voltage control, enable shuttling of many ions in parallel.
Furthermore, SDQC adopts an unidirectional shuttling path to prevent deadlock \cite{delaney_scalable_2024}.

  \item The distance between entanglement factories and detectors (unit distance) is $375~\mu m$ following the previous work \cite{delaney_scalable_2024}. The distance between processor nodes is two unit distances ($750~\mu m$), as a single entanglement factory and detector belong to each processor node, as shown in Fig.~\ref{fig:SDQC}.

  \item The gate teleportation qubits can be reused by transporting them from detectors to entanglement factories after measurements.

\end{enumerate}

\begin{table*}[h]
    \setlength{\tabcolsep}{12pt}
    \renewcommand{\arraystretch}{2}
    \begin{tabular}{ll}
    \hline\hline\\[-2em]
    Component&Method\\[0.2em]
    \hline
    Individual Addressing&\parbox{10cm}{\raggedright AOD \cite{chen_low-crosstalk_2024, hou_individually_2024}, AOM \cite{lim_design_2025}, MEMS \cite{crain_individual_2014}, Integrated photonics \cite{mehta_integrated_2020, mordini_multi-zone_2024, mehta_integrated_2016}, DOM \cite{shih_reprogrammable_2021}}\\[1em]

    State Detection&\parbox{10cm}{\raggedright PMT \cite{noek_high_2013, burrell_scalable_2010, debnath_demonstration_2016}, Integrated photodetector \cite{eltony_transparent_2013}, SNSPD \cite{crain_high-speed_2019, todaro_state_2021}, EM-CCD \cite{burrell_scalable_2010}, sCMOS \cite{sotirova_high-fidelity_2024}, shelving \cite{edmunds_scalable_2021, zhukas_high-fidelity_2021}}\\[1em]

    Shuttling Operation&\parbox{10cm}{\raggedright Linear transportation \cite{delaney_scalable_2024, clark_characterization_2023} Junction rotation \cite{delaney_scalable_2024, burton_transport_2023}, Split \cite{pino_demonstration_2021}, Merge \cite{pino_demonstration_2021}, Physical swap \cite{pino_demonstration_2021}}\\[1em]

    Trap&\parbox{10cm}{\raggedright Segmented DC electrode \cite{pino_demonstration_2021}, Grid-structured surface trap \cite{delaney_scalable_2024}, Two-dimensional micro penning trap \cite{jain_penning_2024}}\\[1em]

    Electrical Source&\parbox{10cm}{\raggedright Co-wired electrode \cite{delaney_scalable_2024}, DC source switching \cite{malinowski_how_2023}}\\
    
    \hline\hline
    \end{tabular}
    \caption{Component-wise demonstrations for the assumptions of SDQC.}
    \label{table:ComponentWiseDemonstration}
\end{table*}

\section{Qubit mapping of superdense color code for SDQC and Photonic DQC}
\label{appendix:PhysicalQubitMappingforLogicalQubit}

\begin{table*}
    \setlength{\tabcolsep}{6pt}
    \renewcommand{\arraystretch}{1.5}
    \begin{tabular}{clcccccccc}
    \hline\hline \\[-1.5em]
         \multirow{2}{*}{Code Distance}&\multirow{2}{*}{Qubit Type}&\multicolumn{6}{c}{Chain}&\multirow{2}{*}{Subtotal}&\multirow{2}{*}{Total}  \\
         &&1&2&3&4&5&6&& \\[0.2em] \hline
         
         &Data&7&&&&&&7& \\
         3&Non-segmented syndrome&6&&&&&&6&13 \\
         &Segmented syndrome&0&&&&&&0& \\ \hline
         
         &Data&19&&&&&&19& \\
         5&Non-segmented syndrome&18&&&&&&18&37 \\
         &Segmented syndrome&0&&&&&&0& \\ \hline
         
         &Data&15&22&&&&&37& \\
         7&Non-segmented syndrome&10&16&&&&&26&73 \\
         &Segmented syndrome&5&5&&&&&10& \\ \hline
         
         &Data&15&24&22&&&&61& \\
         9&Non-segmented syndrome&10&12&16&&&&38&121 \\
         &Segmented syndrome&5&11&6&&&&22& \\ \hline
         
         &Data&15&25&29&22&&&91& \\
         11&Non-segmented syndrome&10&12&14&16&&&52&181 \\
         &Segmented syndrome&5&13&14&6&&&38& \\ \hline
         
         &Data&15&25&17&19&29&22&127& \\
         13&Non-segmented syndrome&10&12&0&0&14&16&52&253 \\
         &Segmented syndrome&5&13&17&18&15&6&74& \\
    \hline\hline
    \end{tabular}
    \caption{Qubit mapping on the ion chains to implement superdense color code for SDQC and Photonic DQC.}
    \label{table:QubitMapping}
\end{table*}

The number of physical qubits (data and syndrome extraction) per logical qubit in Superdense color code $n_\mathrm{ph}$ is given as follows \cite{gidney_new_2023}:
\begin{align}
    n_\mathrm{ph}=\frac{3d_\mathrm{code}^2-1}{2}
    \begin{cases}
        \text{Data qubits: }n_d=\frac{3d_\mathrm{code}^2+1}{4}\\
        \text{Syndrome extraction qubits: }n_a=\frac{3(d_\mathrm{code}^2-1)}{4}
    \end{cases}
\label{equation:NumberofQubits}
\end{align}

Our qubit mapping strategy to implement the superdense color code in DQC architectures (SDQC and Photonic DQC) is to cluster a logical qubit's physical qubits within a single node as much as possible. For SDQC, these nodes are then placed adjacent. 
This clustered mapping minimizes shuttling costs during syndrome measurement while following the constraints of maximum chain capacity.
Table~\ref{table:QubitMapping} shows the details of qubit mapping for the superdense color code, illustrating a maximum chain of 58 physical qubits for $d_\mathrm{code}=13$.

A key advantage of the clustered mapping approach is that most two-qubit gates for syndrome extraction (between data and syndrome extraction qubits) execute within the same processor node.
This is possible because the stabilizer's connectivity topology is localized, requiring syndrome extraction qubits to connect only within their own stabilizer face. In contrast, QCCD architecture projects the color code topology onto its grid-structured trap geometry. As a result, its syndrome extraction qubits should shuttle to adjacent operation zones to realize QEC cycles (Fig.~\ref{fig:ComparisonArchitectures}(a)).

\section{Detailed operation sequences}
\label{appendix:DetailedOperationSequences}

\begin{figure*}
  \includegraphics[width=0.5\paperwidth]{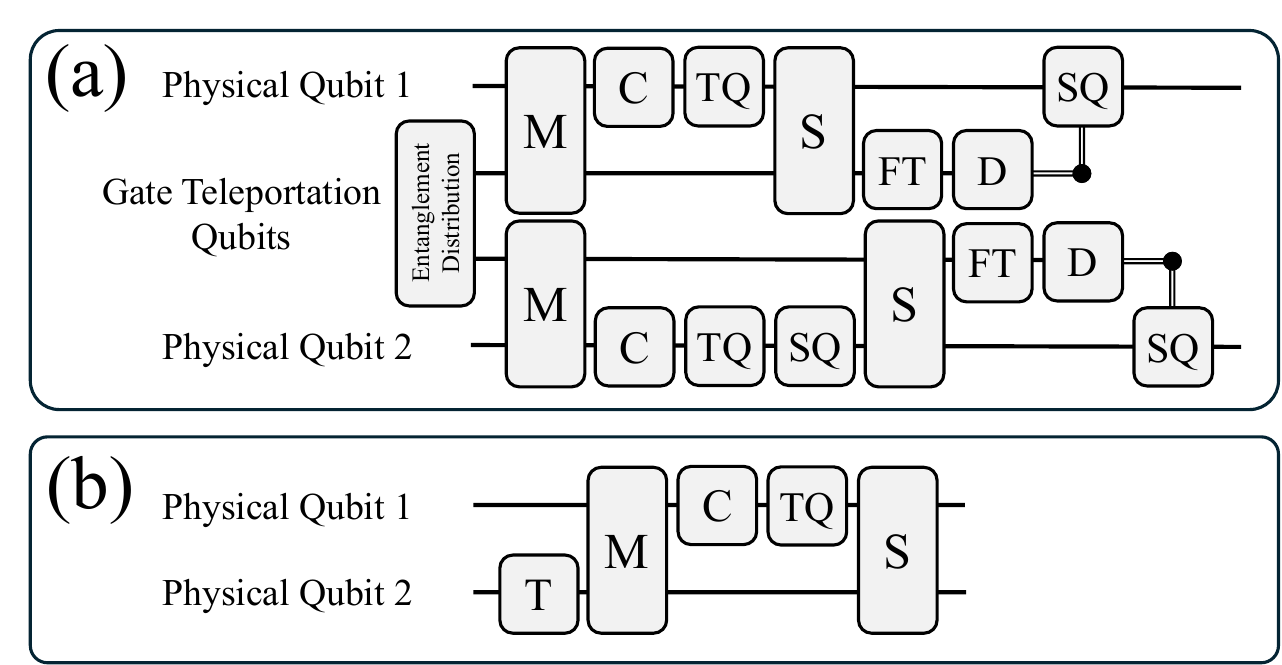}
  \caption{
  Detailed operation sequence to implement a remote physical two-qubit gate for (a) DQC architectures and (b) QCCD.
  (a) DQCs (Photonic DQC and SDQC) employ a gate teleportation protocol to implement a non-local gate. 
  Photonic DQC utilizes photonic entangling for entangling distribution, while SDQC utilizes shuttling operations.
  (b) QCCD employs direct shuttling with a physical qubit to interact with the target physical qubit.
  TQ : Two-qubit Gate, SQ : Single-qubit Gate, T : Adiabatic Transportation, FT : Fast Transportation, M : Merge, S : Split, D : Detection, C : Cooling.}\label{fig:RemoteTwoQubitGateSequence}
\end{figure*}

\begin{figure*}
  \includegraphics[width=0.7\paperwidth]{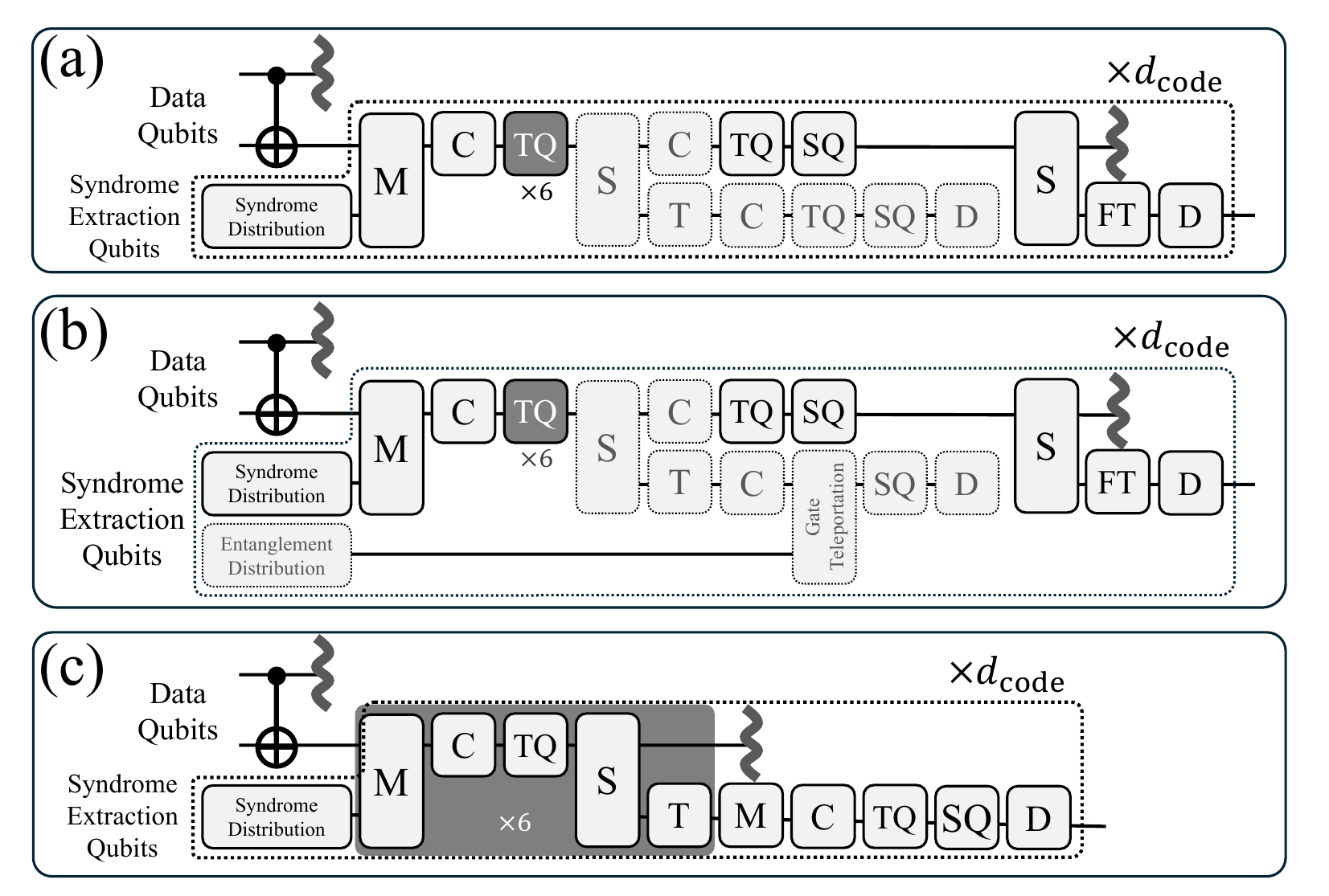}
  \caption{
  Detailed operation sequence to implement syndrome extraction for (a) SDQC, (b) Photonic DQC, and (c) QCCD.
  Syndrome extraction is repeated for $d_\mathrm{code}$ rounds, and segmented stabilizer (dashed gray blocks) are performed for $d_\mathrm{code} \geq 7$.
  The notation follows the same conventions as in Fig.~\ref{fig:RemoteTwoQubitGateSequence}.
  }\label{fig:QECImplemetation}
\end{figure*}

Figure~\ref{fig:RemoteTwoQubitGateSequence} and ~\ref{fig:QECImplemetation} illustrate the detailed operation sequences for physical remote two-qubit gate and syndrome extraction in each architecture.
Actual operations on trapped ion systems include shuttling operations such as split, transportation (stable and fast), and merge, as denoted in the figures.
Whenever a merge or split operation is performed, sympathetic cooling is applied to the ion chain before executing two-qubit gates to suppress motional heating and reduce the error rate.

For physical remote two-qubit gates, DQC architectures employ gate teleportation protocol to achieve long-range connectivity.
On the other hand, QCCD achieves long-range connectivity by data qubit shuttling to reconfigure its connectivity.
For a logical CNOT gate, the transversal gates are implemented by parallel physical two-qubit gates acting across different logical code blocks.

Following a transversal CNOT, syndrome extraction qubits prepared in Bell states are transported and merged with the data qubits to extract the syndromes, and subsequently separated from the processor node for detection.
Syndrome extraction is repeated for $d_\mathrm{code}$ rounds, and for $d_\mathrm{code} \geq 7$, additional operations for segmented stabilizers are included, as indicated by gray dashed blocks in Fig.~\ref{fig:QECImplemetation}.
Due to the architectural similarity between SDQC and Photonic DQC, the syndrome extraction sequences of the two architectures are identical, except for the Bell measurement process of syndrome extraction qubits associated with segmented stabilizers. 
In Photonic DQC, an additional gate teleportation is required to connect the syndrome extraction qubits of segmented stabilizers located in different ELUs, whereas in SDQC, the corresponding syndrome extraction qubits are shuttled and collected into the same detector, eliminating the need for remote gate teleportation.

\section{Time cost calculation}
\label{appendix:TimeCostCalculation}

\subsection{Remote physical two-qubit gate}
\label{appendix:TimeCostCalculationRemotePhysicalTwoQubitGate}

The time cost calculation for the remote physical two-qubit gate is based on the operation sequence illustrated in Figure~\ref{fig:RemoteTwoQubitGateSequence}.
For each architecture, we determine the total execution time by identifying the longest path among all parallelizable operations and summing the base time costs of the operations on that path (see Sec.~\ref{subsection:TimeCostSetup}).
Table~\ref{table:RemotePhysicalTQGate} shows the specific operations and shuttling included in the calculation.
DQC architectures require preparatory entanglement distribution, which can be asynchronous and hidden by pipelining, denoted as numbers in parentheses in the Table~\ref{table:RemotePhysicalTQGate}.

\begin{table*}[h]
    \setlength{\tabcolsep}{3pt}
    \renewcommand{\arraystretch}{1.5}
    \begin{tabular}{lcccc ccccc c}
    \hline\hline
         \multirow{2}{*}{Architecture}&\multicolumn{8}{c}{Operations} & \multicolumn{2}{c}{Transportation}\\
         & \parbox{1.7cm}{Two-qubit gate} & \parbox{2cm}{Single-qubit gate} & Cooling & Merge 
         & Split & Detection & Swap & \parbox{2cm}{Photonic\\entangling} 
         & Stable & Fast \\[1em]
         \hline
         
         SDQC         
         & 1 (1)  & 1 (1) & 1 & 1 
         & 1 (1) & 1 & 0 & 0         
         & 0 (3+$\bar{l}_{\text{SDQC}}$) & 2 \\
         
         Photonic DQC 
         & 1      & 1     & 1 & 1 
         & 1 & 1 & 0 & 0 (1)
         & 0 (3) & 2 \\
         
         QCCD         
         & 1 & 0 & 1 & $\bar{n}_{\text{swap}}$
         & $\bar{n}_{\text{swap}}$ & 0 & $\bar{n}_{\text{swap}}$ & 0 
         & $\bar{l}_{\text{QCCD}}$ & 0 \\
    \hline\hline
    \end{tabular}
    \caption{
       Operations contributing to the latency of a remote two-qubit gate.
       Numbers in parentheses indicate extra operations required for entanglement distribution.
       Unit distance for shuttling is $375~\mu m$ for all architectures.
       }
    \label{table:RemotePhysicalTQGate}
\end{table*}

In shuttling-based architectures, the time cost for shuttling-related operations depend on the system scale.
In SDQC, the average shuttling distance of gate teleportation qubits represented as $\bar{l}_{\text{SDQC}}$ increases as the qubit number and code distance grows.
In SDQC, the average shuttling distance for gate teleportation qubits, denoted $\bar{l}_{\text{SDQC}}$, increases with both the number of logical qubits and the code distance.
In QCCD, the average distance of data qubit shuttling $\bar{l}_{\text{QCCD}}$ and the average number of swap operations $\bar{n}_{\text{swap}}$ scale as the square root of the system size \cite{webber_efficient_2020}.
The detailed derivations is given as follows:

\begin{align}
    \bar{l}_{\text{SDQC}} &= 2n_c\sum_{k=1}^{n_\mathrm{L}-1}{\frac{k(n_\mathrm{L}-k)}{_{n_\mathrm{L}}C_2}} =\frac{2n_c(n_\mathrm{L}+1)}{3} 
    \label{equation:SDQCMeanDistance}\\
    \bar{l}_{\text{QCCD}} &=  1.3\sqrt{n_\mathrm{ph}n_\mathrm{L}}+2 = 1.3\sqrt{\frac{3d_\mathrm{code}^2-1}{2}n_\mathrm{L}}+2\\
    \bar{n}_{\text{swap}} &= 0.23\sqrt{n_\mathrm{ph}n_\mathrm{L}}+0.1 = 0.23\sqrt{\frac{3d_\mathrm{code}^2-1}{2}n_\mathrm{L}}+0.1 \label{equation:QCCDRouting}
\end{align}

\noindent where $n_\mathrm{L}$ is the number of logical qubits and $n_c$ is the number of ion chains per logical qubit (see Appendix~\ref{appendix:PhysicalQubitMappingforLogicalQubit}).

\subsection{Detailed operation scheduling for logical operations}
\label{appendix:Pipelining}
\begin{figure*}
  \includegraphics[width=0.7\paperwidth]{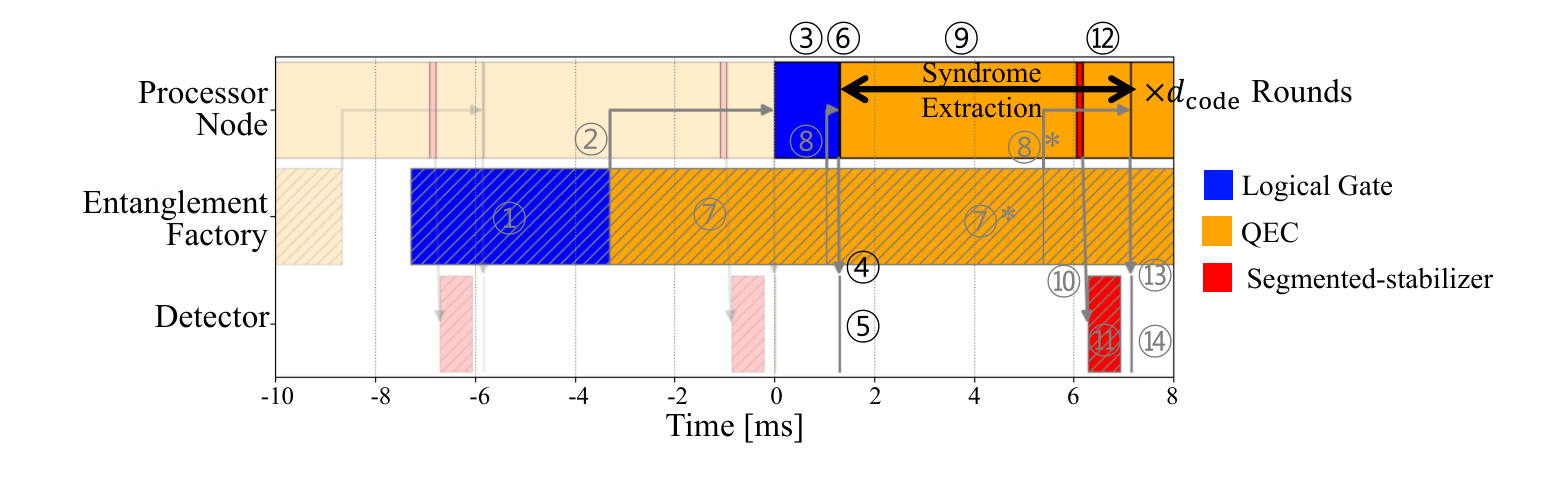}
  \caption{Pipelining of Photonic DQC for $d_\mathrm{code}=13$.
  Photonic DQC for larger $d_\mathrm{code}$ than 7 requires additional gate teleportation using photonic entangling to connect with other modules for Bell measurement of segmented stabilizers.
  }{\label{fig:PhotonicDQCPipelining}}
\end{figure*}
The goal of the operation scheduling is to minimize the overall time cost of a logical operation consisting of transversal CNOT gates and $d_\mathrm{code}$ rounds of syndrome extraction
cycles.
This is achieved by reducing sequential latency and enhancing parallelism via pipelining, which ultimately shortens the logical clock time.

SDQC can ideally eliminate the latency associated with entanglement distribution via pipelining (as shown in Fig.~\ref{fig:SDQCPipelining}(b)).
This is enabled by SDQC's multi-site implementation and sufficient entanglement throughput, as described in Sections \ref{subsection:Piplining} and \ref{subsubsection:TimeLogicalTQ}.

On the other hand, QCCD is difficult to adopt pipelining between transversal CNOT gates and syndrome extraction cycles.
Different from SDQC, which uses separate entanglement distribution, QCCD must directly move its data qubits for transversal CNOT gates. As both transversal CNOT gates and syndrome extraction cycles utilize the data qubits, it is difficult for QCCD to execute them in parallel with pipelining.

Finally, Figure~\ref{fig:PhotonicDQCPipelining} shows the pipelining for Photonic DQC.
The overall scheduling is conceptually similar to that of SDQC.
Similar to SDQC, Photonic DQC can also parallely execute a logical CNOT gate and syndrome extraction cycle with pipelining.

\subsection{Syndrome extraction}
\label{appendix:SyndromeExtraction}

\begin{table*}[h]
    \setlength{\tabcolsep}{3pt}
    \renewcommand{\arraystretch}{1.5}
    \begin{tabular}{lcccccccc}
    \hline\hline
         \multirow{2}{*}{Architecture}&\multicolumn{6}{c}{Operations} & \multicolumn{2}{c}{Transportation}\\
         & \parbox{1.7cm}{Two-qubit gate} & \parbox{2cm}{Single-qubit gate} & Cooling & Merge 
         & Split & Detection 
         & Stable & Fast \\[1em]
         \hline
         
         SDQC         
         & 7 & 1 & 1 (1) & 1 
         & 1 (1) & 1 
         & 0 & 2 \\
         
         Photonic DQC 
         & 7 & 1 & 1 (1) & 1 
         & 1 (1) & 1 
         & 0 & 2 \\
         
         QCCD         
         & 7 & 1 & 7 & 7
         & 6 & 1
         & 20 & 0 \\
    \hline\hline
    \end{tabular}
    \caption{
       Operations contributing to the latency of a single round of syndrome extraction.
       Numbers in parentheses indicate extra operations required when $d_{\text{code}}\geq 7$.
       Unit distance for shuttling is $375~\mu m$ for all architectures.
       }
    \label{table:LogicalTQGate}
\end{table*}

From the operation sequence and pipelined schedule (see Appendix~\ref{appendix:DetailedOperationSequences},\ref{appendix:Pipelining}), the total latency of a single round of syndrome extraction is obtained by identifying the critical path.
Table~\ref{table:LogicalTQGate} lists the operations that appear on this path for each architecture.
For code distances $d_{\text{code}}\geq 7$, segmented stabilizers require a modified sequence, which increases the length of each syndrome-extraction round.

\clearpage

\section{Error rate calculation for remote physical two-qubit gates}
\label{appendix:PhysicalErrorRate}

The remote physical two-qubit gate error rate $p_\mathrm{trans}$ can be calculated based on the number of operations, total execution time, and the number of traversed junctions during ion shuttling. 
The base error rates specified in Table~\ref{table:BaseErrorRate} are denoted as the single-qubit gate error rate $p_\mathrm{SQ}$, the two-qubit gate error rate $p_\mathrm{TQ}$, the measurement error rate $p_\mathrm{meas}$, the photonic entangling error rate $p_\mathrm{pe}$, the error per junction traversal $p_\mathrm{j}$, and the idle error rate per 1 ms $p_\mathrm{idle}$, respectively, in order.
We model the remote physical error rate as a linear sum of their individual contributions:

\begin{align}
    p_\mathrm{trans}= p_\mathrm{o}+\bar n_j p_j+ \Sigma T_\mathrm{decohere} p_\text{Idle}.
    \label{equation:PhysicalErrorModel}
\end{align}

Here, $p_o$ represents the total error from active operations, including the contributions of gate operations, detections, and photonic entanglement generations.
The second term denotes the total shuttling error, where $\bar{n}_j$ is the average number of junctions traversed during shuttling.
The final term models the cumulative decoherence error, determined by total decoherence exposure $\Sigma T_\mathrm{decohere}$, which is the effective decoherence exposure (in units of ms) of all qubits engaged in the process.

All these parameters highly depend on the detailed architectural configuration, QEC configuration, and system scale.
We calculate $p_{o}$, $\bar n_j$, and $\Sigma T_\mathrm{decohere}$ of each architecture as follows:

\begin{align}
\text{SDQC}&:~
    \begin{cases}
        p_o= 3\times p_\mathrm{TQ} + 2\times p_\mathrm{meas} +  3\times p_\mathrm{SQ},\\
        \\
        \bar n_j=(4+\bar{l}_\mathrm{SDQC})+2,\\
        \\
        \begin{aligned}
        \Sigma T_\mathrm{decohere}=2 T_\mathrm{ED}+4T_\mathrm{Remote TQ},
        \end{aligned}
    \end{cases}
\label{equation:SDQCArchitectureParameters} \\
\text{QCCD}&:~
    \begin{cases}
        p_o= p_\mathrm{TQ},\\
        \\
        \bar{n}_j = 2\times(0.4\sqrt{n_\mathrm{ph}n_\mathrm{L}}+2),\\
        \\
        \Sigma T_\mathrm{decohere}=2T_\mathrm{Remote TQ},
    \end{cases}
\label{equation:QCCDArchitectureParameters} \\
\text{Photonic DQC}&:~
    \begin{cases}
        p_o=p_\mathrm{pe} + 2\times p_\mathrm{TQ}+ 2\times p_\mathrm{meas} +3\times p_\mathrm{SQ},\\
        \\
        \bar n_j=(4)+2,\\
        \\
        \begin{aligned}
        \Sigma T_\mathrm{decohere}=2 T_\mathrm{ED}+4T_\mathrm{Remote TQ},
        \end{aligned}
    \end{cases}
\label{equation:PhotonicDQCArchitectureParameters}
\end{align}
\noindent where $T_\mathrm{ED}$ and $T_\mathrm{Remote TQ}$ represent entanglement distribution time and remote two-qubit gate time, respectively.

We model $p_o$ of SDQC and Photonic DQC following the operation sequence of their remote two-qubit gate (Fig.~\ref{fig:SDQC}(b)).
Step~1 in Fig.~\ref{fig:SDQC}(b) is implemented with a CNOT gate for SDQC, and with a photonic entanglement for Photonic DQC, respectively.
We model $p_o$ of QCCD as a two-qubit gate error rate because the operation sequence for its remote two-qubit gate is implemented with a single CNOT gate after the ion shuttling.

Second, we derive $\bar{n}_j$ of three architectures following their overall structure and shuttling path of their qubits.
For SDQC and Photonic DQC, each term inside the equations represents the number of traversed junctions for Steps~2 and 4 in Fig.~\ref{fig:SDQC}(b), respectively. For SDQC, Step~2 involves moving two qubits entangled in Bell pairs. One qubit moves to the processor node directly above the entanglement factory, traversing two junctions. The other qubit also traverses these two junctions and then moves an additional $\bar{l}_\mathrm{SDQC}$ junctions on average to its destination node, resulting in $2+d$ traversals in total. Here, $\bar{l}_\mathrm{SDQC}$ represents the average distance between nodes derived from Eq.~\eqref{equation:SDQCMeanDistance}. However, this error can be mitigated by spare qubits, and hence does not contribute to the overall error rate.
For the second term (Step~4) of SDQC, the two qubits are moved to the detector directly below each node, resulting in two junction traversals in total. For QCCD, we utilize the formulation from the previous work \cite{webber_efficient_2020}. $n_\mathrm{L}$ is the number of logical qubits and $n_\mathrm{ph}$ is the number of physical qubits per logical qubit for the superdense color code derived in Eq.~\eqref{equation:NumberofQubits}. For Photonic DQC, the long-distance shuttling in Step~2 (i.e., traversing $\bar{l}_\mathrm{SDQC}$ junctions in SDQC) is eliminated thanks to the photonic entanglement. This results in four junction traversals for Step~2 and two junction traversals for Step~4. Note that for both SDQC and Photonic DQC, the term for Step~2 (i.e., first term) can be eliminated if we utilize spare qubits to mitigate the shuttling error.

Finally, we derive the total decoherence exposure of SDQC and Photonic DQC following Fig.~\ref{fig:SDQC}(b).
For SDQC and Photonic DQC, this process involves two data qubits and two gate teleportation qubits.
The decoherence exposure is modeled differently for each qubit type: the exposure of gate teleportation qubits is the sum of Steps~1-6 (\textit{i.e.}, $T_\mathrm{ED} + T_\mathrm{remoteTQ}$), whereas the time for data qubits is modeled only as the sum of Steps~3-6 (\textit{i.e.}, $T_\mathrm{remoteTQ}$).
As the data qubits perform syndrome extraction during Steps~1-2, their decoherence error during Steps~1-2 is accounted in the syndrome extraction process.
In contrast, for QCCD, this process involves two data qubits, and their decoherence exposures are the sum of the shuttling duration and the physical CNOT gate duration (\textit{i.e.}, $T_\mathrm{remoteTQ}$).

\section{Logical error rate calculation}
\label{appendix:LogicalErrorRateCalculation}
\subsection{Noise model for QEC simulation circuit}
To conduct QEC simulations, we inject noise into these operational sequences based on a circuit-level noise model.
We denote the single-qubit and two-qubit depolarizing noise channels with error rate $p$ as $\mathrm{DEP}_1(p)$ and $\mathrm{DEP}_2(p)$, respectively.
Then the noise model, characterized by the transversal gate error rate $p_\mathrm{trans}$ and the operation improvement factor for syndrome extraction $\lambda_\mathrm{SE}$, is given as follows:
\begin{itemize}
    \item A transversal CNOT gate is replaced with a set of channels $\mathrm{DEP}_1(4p_\mathrm{trans}/5)$ that act on all data qubits (see Section~\ref{subsubsec:logical_error_rate_setup} for details).
    \item Every Hadamard gate for Bell initialization or measurement is followed by $\mathrm{DEP}_1(p_\mathrm{SQ} / \lambda_\mathrm{SE})$.
    \item Every physical CNOT gate is followed by $\mathrm{DEP}_2(p_\mathrm{TQ}/\lambda_\mathrm{SE})$. Exceptionally, in Photonic DQC, CNOT gates used in the Bell measurements for segmented stabilizers are preceded by $\mathrm{DEP}_1((p_\mathrm{pe}+p_\mathrm{TQ}+p_\mathrm{SQ}+p_\mathrm{meas})/\lambda_\mathrm{SE})$, which accounts for the noise from gate teleportation using photonic entangling operations.
    \item (Only for QCCD) Every physical CNOT gate between data and syndrome extraction qubits is preceded by $\mathrm{DEP}_1 (p_\mathrm{j})$ on the syndrome extraction qubit, as performing the gate requires a shuttling operation of the syndrome extraction qubit passing over one junction, which introduces an ion loss error.
    \item Every syndrome extraction qubit measurement outcome is flipped with probability $p_\mathrm{meas}/\lambda_\mathrm{SE}$.
    \item When syndrome extraction qubits are shuttled for Bell measurements, they experience noise modeled as $\mathrm{DEP}1(n_\mathrm{j} p_\mathrm{j} / \lambda_\mathrm{SE})$, where $n_\mathrm{j}$ is the number of junction crossings during shuttling. We take $n_\mathrm{j} = 2$ ($n_\mathrm{j} = 3$) for each segment of the segmented stabilizers in SDQC (Photonic DQC), and $n_\mathrm{j} = 1$ in all other cases. In general, a single junction crossing is required to reach the nearest operation zone or detector through shuttling, as illustrated in Fig.~\ref{fig:SDQC} and Fig.~\ref{fig:ComparisonArchitectures}. Segmented stabilizers, however, incur additional crossings, either when shuttling to the next detector in SDQC or when shuttling for gate teleportation in Photonic DQC.
    \item Qubits that are idling, shuttling, or transported are subjected to idling noise modeled as $\mathrm{DEP}_1(p_\mathrm{idle} T / \lambda_\mathrm{SE})$, where $T$ is the corresponding required time based on Table~\ref{table:UnitOperationTime}. For example, for data qubits that idle during CNOT gates between syndrome and data qubits, we set $T = \mathrm{Max}(13.33N-54, 100)$ for SDQC and Photonic DQC as specified in Table~\ref{table:UnitOperationTime}, where $N$ is the maximum number of physical qubits in a node calculated from Table~\ref{table:QubitMapping}. For QCCD, we assume $T = 749.8$ $\mu$s, combining shuttling ($2 \times 46.9$ $\mu$s), merge (128 $\mu$s), cooling (300 $\mu$s), CNOT (100 $\mu$s), and split (128 $\mu$s) operations. $T$ is computed in a similar way for shuttling or transported qubits.
\end{itemize}

\subsection{QEC simulation results}
\label{appendix:QECSimulationResult}

QEC simulations for each architecture are performed based on the simulation setup described in Sec.~\ref{subsubsection:BaselineArchitecture}, under the architectural setup shown in Sec.~\ref{subsubsec:logical_error_rate_setup} and Appendix \ref{appendix:DetailedOperationSequences}.

Figures~\ref{fig:QECSimulationResult_SDQC}--\ref{fig:QECSimulationResult_PDQC} present the computed logical error rates for SDQC, QCCD, and Photonic DQC, respectively, with varying code distances ($d_\mathrm{code}$), transversal gate error rates ($p_\mathrm{trans}$), and operation improvement factors for syndrome extractions ($\lambda_\mathrm{SE}$).
Each color in the figure corresponds to the different $\lambda_\mathrm{SE}$ values used in our simulation.
Each curve represents the fitted line for the data of each $\lambda_\mathrm{SE}$ value. 
For curve fitting, we utilize the ansatz presented in Eq.~\eqref{equation:LogicalErrorRateModel}.
Their parameter and $R^2$ values are summarized in Table~\ref{table:LogicalErrorRateFittingParameters}, together with the crossover error rates $p^*_\mathrm{trans}$ at $\lambda_\mathrm{SE}=1$, verifying that the fits are reasonably accurate ($R^2 > 0.97$) in all cases except for Photonic DQC with $d_\mathrm{code} \geq 7$.

We attribute this exception to the following: In Photonic DQC with $d \geq 7$, segmented stabilizer measurements rely on photonic entangling operations that are orders of magnitude noisier than the other operations (see Table~\ref{table:BaseErrorRate}).
Consequently, the logical error rate remains strongly dependent on $\lambda_\mathrm{SE}$ even at relatively large $p_\mathrm{trans}$ ($\gtrapprox 10^{-2}$), so the simulated data do not clearly enter the ``transversal-gate-dominated'' regime (i.e., $p_\mathrm{trans} > p^*_\mathrm{trans}$) where the curves for different $\lambda_\mathrm{SE}$ converge, reducing the quality of the fit.
Nevertheless, we estimate that our comparison of SDQC with Photonic DQC is still conservative, as the fitted curves for Photonic DQC in Fig.~\ref{fig:QECSimulationResult_PDQC} generally predict lower logical error rates than those observed in the raw data, particularly around $p_\mathrm{trans} \approx 3 \times 10^{-2}$ in Fig.~\ref{fig:ErrorRate_d13}.

\begin{figure*}
\includegraphics[width=0.75\textwidth, keepaspectratio]{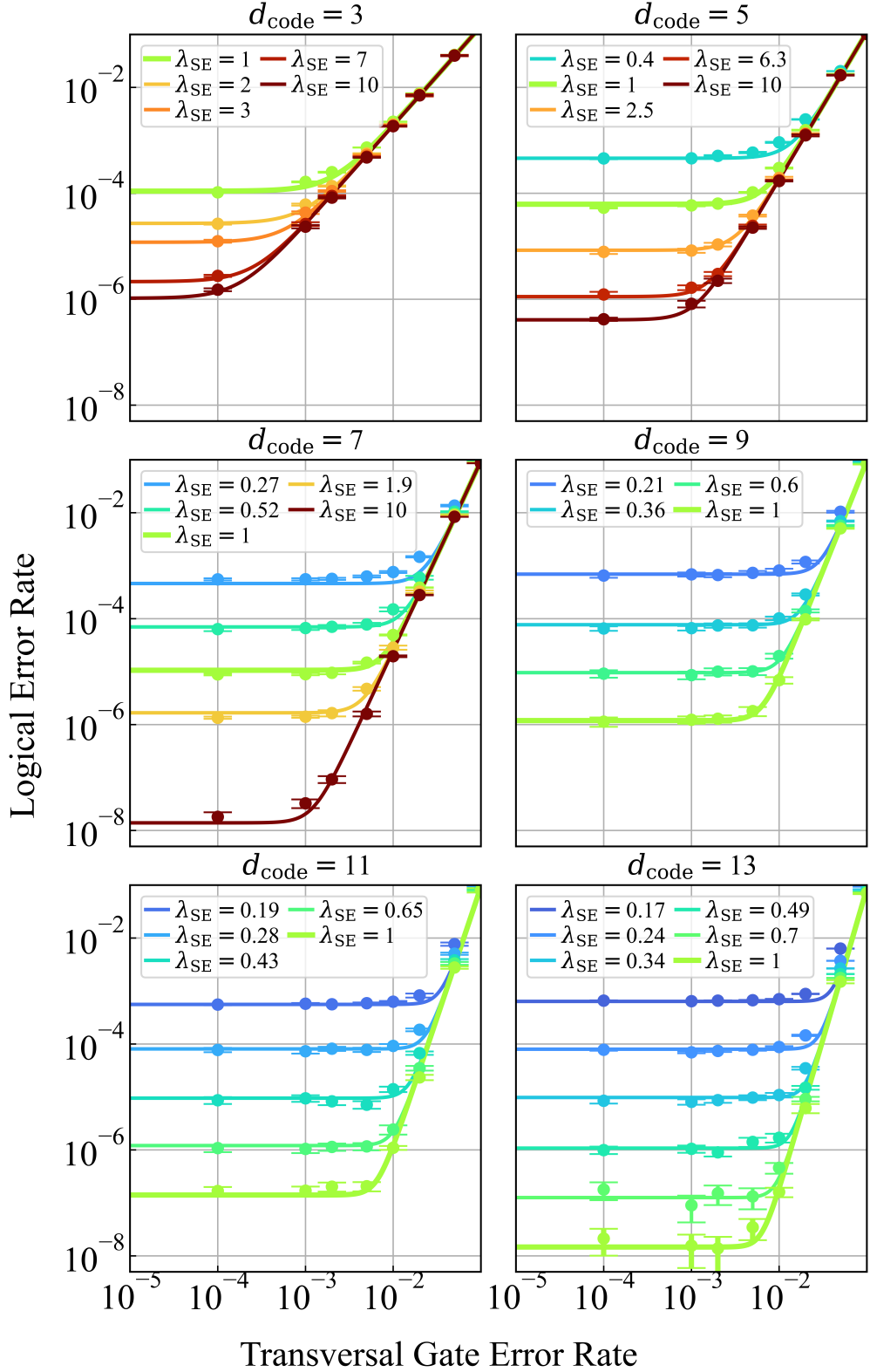}
\caption{QEC simulation results for SDQC, based on the setup in Sec.~\ref{subsubsec:logical_error_rate_setup}. The logical error rates are plotted as a function of the transversal gate error rate ($p_\mathrm{trans}$) with varying code distances ($d_\mathrm{code}$) and the operation improvement factor for syndrome extraction ($\lambda_\mathrm{SE}$). Error bars represent 1-sigma regions. The solid lines indicate curves fitted to the ansatz of Eq.~\eqref{equation:LogicalErrorRateModel}. The cases of $\lambda_\mathrm{SE}=1$ are highlighted as thick lines.}
\label{fig:QECSimulationResult_SDQC}
\end{figure*}

\begin{figure}
\includegraphics[width=0.75\textwidth,keepaspectratio]{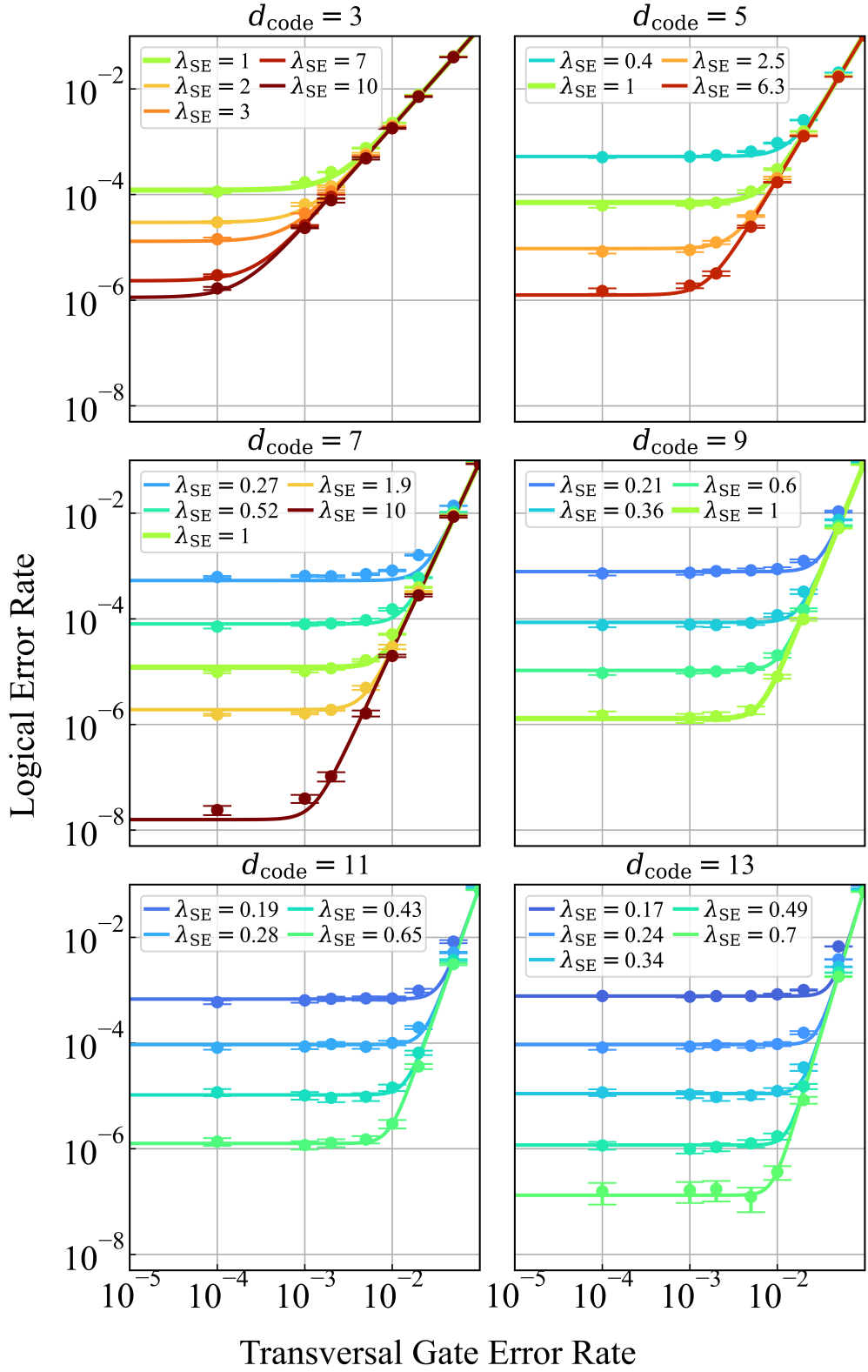}
\caption{QEC simulation results for QCCD.}
\label{fig:QECSimulationResult_QCCD}
\end{figure}

\begin{figure}
\includegraphics[width=0.75\textwidth,keepaspectratio]{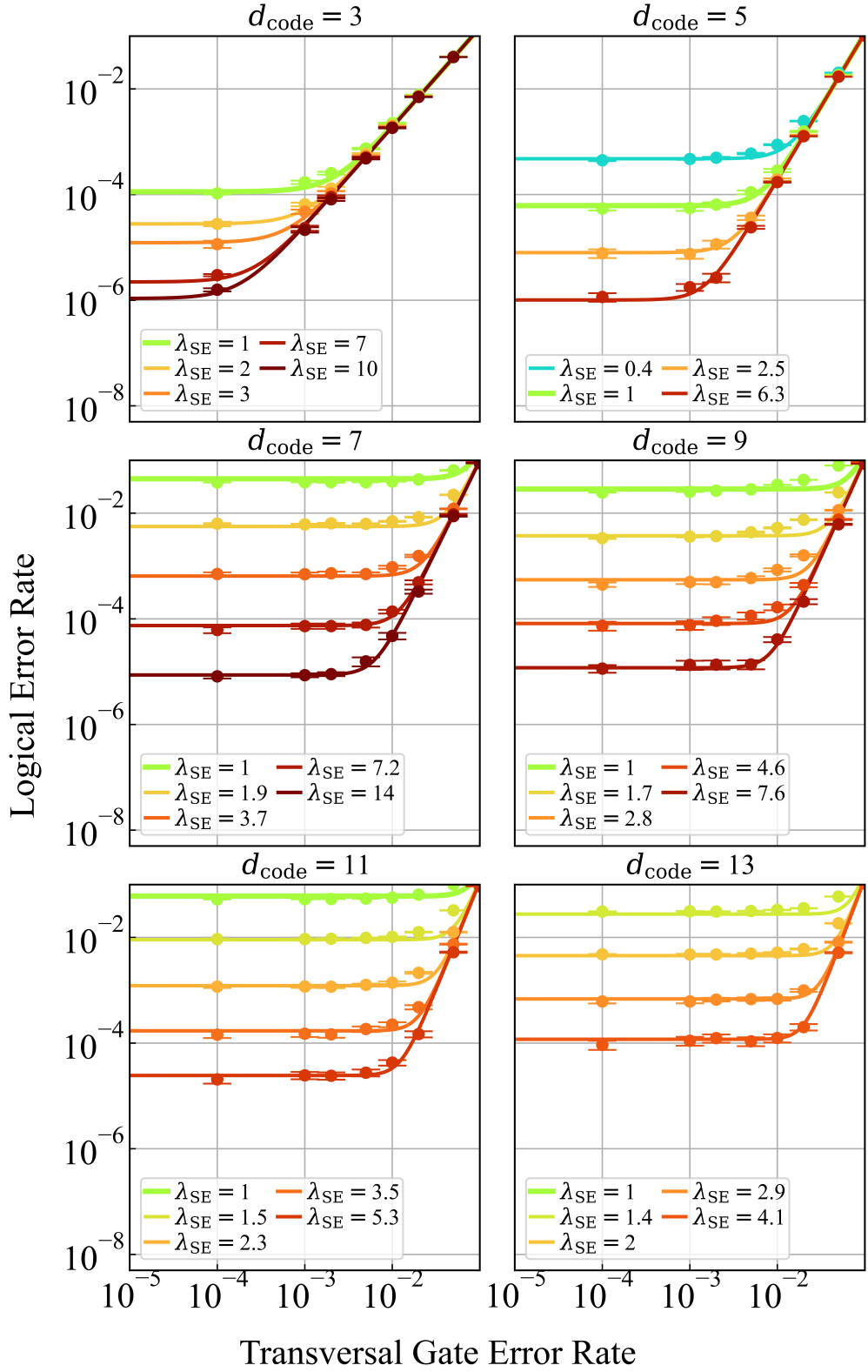}
\caption{QEC simulation results for Photonic DQC.}
\label{fig:QECSimulationResult_PDQC}
\end{figure}

\begin{table*}
    \renewcommand{\arraystretch}{1.5}
    \centering
    \begin{tabular}{cccccccc}
    \hline\hline\\[-1em]
    \parbox{2cm}{Architectures}&\parbox{2cm}{Code distance}&\parbox{2cm}{A}&\parbox{2cm}{B}&\parbox{2cm}{$\alpha$}&\parbox{2cm}{$\beta$}&$R^2$&$p^*_\mathrm{trans}$ at $\lambda_\mathrm{SE}=1$\\[1em]
    \hline
    \multirow{6}{*}{SDQC}&3&5.29(16)&5.48(20)$\times10^{-5}$&0.624(2)&0.674(9)&0.9977&$2.17\times10^{-3}$\\
    &5&3.40(15)$\times10$&3.10(7)$\times10^{-5}$&0.557(2)&0.436(4)&0.9929& $6.78\times10^{-3}$\\
    &7&1.80(16)$\times10^{2}$&5.29(17)$\times10^{-6}$&0.511(3)&0.412(4)&0.9922& $7.84\times10^{-3}$\\
    &9&6.56(1.23)$\times10^{2}$&5.94(36)$\times10^{-7}$&0.460(6)&0.453(6)&0.9876& $6.54\times10^{-3}$\\
    &11&3.98(89)$\times10^{3}$&7.02(51)$\times10^{-8}$&0.449(7)&0.454(6)&0.9853& $6.65\times10^{-3}$\\
    &13&2.08(51)$\times10^{4}$&7.33(61)$\times10^{-9}$&0.437(6)&0.464(6)&0.9713& $6.43\times10^{-3}$\\
    \hline
    \multirow{6}{*}{QCCD}&3&5.31(20)&6.07(24)$\times10^{-5}$&0.624(3)&0.677(10)&0.9979& $2.29\times10^{-3}$\\
    &5&3.27(32)$\times10$&3.52(11)$\times10^{-5}$&0.553(6)&0.438(7)&0.9926& $6.94\times10^{-3}$\\
    &7&1.76(17)$\times10^{2}$&6.06(20)$\times10^{-6}$&0.510(4)&0.412(4)&0.9926& $8.12\times10^{-3}$\\
    &9&6.00(1.16)$\times10^{2}$&6.47(39)$\times10^{-7}$&0.456(7)&0.456(7)&0.9875& $6.53\times10^{-3}$\\
    &11&2.51(57)$\times10^{3}$&6.95(61)$\times10^{-8}$&0.430(7)&0.465(7)&0.9834& $5.86\times10^{-3}$\\
    &13&1.67(41)$\times10^{4}$&7.37(68)$\times10^{-9}$&0.428(6)&0.472(6)&0.9729& $6.02\times10^{-3}$\\
    \hline
    \multirow{6}{*}{Photonic DQC}&3&5.44(20)&5.63(25)$\times10^{-5}$&0.627(3)&0.673(11)&0.9973& $2.24\times10^{-3}$\\
    &5&3.40(34)$\times10$&3.07(10)$\times10^{-5}$&0.556(6)&0.447(7)&0.9910& $6.70\times10^{-3}$\\
    &7&1.22(24)$\times10^{2}$&9.81(101)$\times10^{-3}$&0.483(9)&0.401(4)&0.9773& $6.15\times10^{-2}$\\
    &9&2.44(85)$\times10^{2}$&1.43(11)$\times10^{-2}$&0.409(13)&0.427(7)&0.8787& $7.08\times10^{-2}$\\
    &11&7.16(2.33)$\times10^{2}$&3.03(16)$\times10^{-2}$&0.372(10)&0.426(5)&0.9118& $8.54\times10^{-2}$\\
    &13&1.63(74)$\times10^{3}$&7.62(35)$\times10^{-2}$&0.339(14)&0.391(4)&0.9420& $1.04\times10^{-1}$\\
    \hline\hline
    \end{tabular}
    \caption{Fitted parameters of the logical error rate model [Eq.~\eqref{equation:LogicalErrorRateModel}]. The corresponding values for $R^2$ and the crossover error rates $p^*_\mathrm{trans}$ at $\lambda_\mathrm{SE}=1$ are additionally specified.}
    \label{table:LogicalErrorRateFittingParameters}
\end{table*}

\clearpage

\section{Space cost calculation}
\label{appendix:SpaceCostCalculation}

The space cost in our evaluation indicates the total number of physical qubits required for executing target applications. 
The space cost depends on the architecture, code distance, and configuration of the application (e.g., the number of gates and logical qubits).
The space cost can be categorized into the data qubit count, qubit count for syndrome extraction, and qubit count for gate teleportation.

\begin{align}
    \text{SpaceCost}_\text{SDQC} &= (n_dn_\mathrm{L})+(2n_an_\mathrm{L})+(2(n_d+n_\mathrm{spare}) \times n_{gate/layer})\label{equation:SpaceCost_sdqc}
    \\
    \text{SpaceCost}_\text{QCCD} &= (n_dn_\mathrm{L})+(2n_an_\mathrm{L}) + 0\label{equation:SpaceCost_qccd}
    \\
    \text{SpaceCost}_\text{PhotonicDQC} &= (n_dn_\mathrm{L})+(2n_an_\mathrm{L}+2n_sn_\mathrm{L})+(2n_d \times n_{gate/layer})\label{equation:SpaceCost_pdqc}
\end{align}\label{equation:SpaceCost}

Equations \ref{equation:SpaceCost_sdqc}, \ref{equation:SpaceCost_qccd}, and \ref{equation:SpaceCost_pdqc} represent the space cost of SDQC, QCCD, and Photonic DQC.
Each term inside the equations corresponds to the data qubit count, the qubit count for syndrome extraction, and the qubit count for gate teleportation, respectively.
In the equation, $n_d$, $n_a$, and $n_s$ represent the number of data qubits, total syndrome extraction qubits, and segmented syndrome extraction qubits per logical qubit in the superdense color code shown in Fig.~\ref{fig:QECMapping}.
We summarize the exact values of $n_d$, $n_a$, and $n_s$ in Table~\ref{table:QubitMapping}.
$n_\mathrm{L}$ is the number of logical qubits, and $n_{gate/layer}$ represents the maximum two-qubit gate number per layer of the target application.
For Fermi-Hubbard model, the $n_\mathrm{L}$ and $n_{gate/layer}$ are \num{132} logical qubits and \num{42} two-qubit gates, respectively.
For ECDLP, the $n_\mathrm{L}$ and $n_{gate/layer}$ are \num{2871} logical qubits and \num{343} two-qubit gates, respectively.

The number of spare qubits $n_\mathrm{spare}$ is calculated to make the ion loss probability during entanglement distribution below 1~\% of the transversal gate error rate.

\begin{align}
    p_\mathrm{loss/gate} &= 1-\sum^{n_\mathrm{spare}}_{n=0}p_\mathrm{loss/pair}^n(1-p_\mathrm{loss/pair})^{n_d+n_\mathrm{spare}-n}\binom{n_d+n_\mathrm{spare}}{n}\label{equation:lossprob} \\
    p_\mathrm{loss/pair} &= 1-(1-\varepsilon_j)^{\bar{n}_j}\label{equation:lossprob2}
\end{align}

Equation \eqref{equation:lossprob} represents the ion loss probability of losing more than $n_\mathrm{spare}$ qubits from the total pool of $n_d + n_\mathrm{spare}$ qubits during a transversal gate operation.
This probability depends on $n_\mathrm{spare}$ and ion loss probability per entanglement pair $p_\mathrm{loss/pair}$.
$p_\mathrm{loss/pair}$ is calculated using Equation \eqref{equation:lossprob2}, which is a function of the shuttling error of moving an ion over a junction $\varepsilon_j$ (i.e., Shuttling in Table~\ref{table:BaseErrorRate}) and the number of traversed junctions during shuttling $\bar{n}_j$.
The junction count $\bar{n}_j$ is derived from Equation \eqref{equation:SDQCArchitectureParameters}.

\end{widetext}
\bibliography{references.bib}
\end{document}